\documentclass[letterpaper]{raa}            % referee version: for submission

\usepackage{graphicx,times}             %for PS/EPS graphics inclusion, new
\usepackage{multirow}
\usepackage{makecell}
\begin{document}
\title{Morphology of the $^{13}$CO(3-2) millimeter emission across the gas disc 
surrounding the triple protostar GG Tau A using ALMA observations
%\footnotetext{$*$ Supported by the National Natural Science Foundation of China.}
}
%   \subtitle{I. Place Your Subtitle Here}

   \volnopage{Vol.0 (200x) No.0, 000--000}      %%preserved for Editor. DOn't remove!
   \setcounter{page}{1}          %%starting page, preserved for Editor. DOn't remove!

   \author{N.T. Phuong \inst{1^\star,2,3}
   \and P.N. Diep \inst{1,2}
   \and A. Dutrey \inst{3}
   \and E. Chapillon \inst{3,4}
   \and P. Darriulat \inst{1}
   \and S. Guilloteau \inst{3}
   \and \mbox{D.T. Hoai \inst{1}}
   \and P.T. Nhung \inst{1}
   \and Y.-W. Tang \inst{5}
   \and N.T. Thao \inst{1}
   \and P. Tuan-Anh \inst{1}
}

   \institute{Department of Astrophysics, Vietnam National Space Center, 
Vietnam Academy of Science and Techonology, 18 Hoang Quoc Viet, Cau Giay, Hanoi, Vietnam; 
   {\it {$^\star$ ntphuong02@vnsc.org.vn}}\\
   \and Graduate University of Science and Technology, Vietnam Academy of Science and Techonology, 18 Hoang Quoc Viet, Cau Giay, Hanoi, Vietnam \\
   \and Laboratoire d'astrophysique de Bordeaux, Universit\'e de Bordeaux, CNRS, B18N, All\'ee Geoffroy Saint-Hilaire, F-33615 Pessac \\
   \and IRAM, 300 rue de la piscine, F-38406 Saint Martin d'H\`eres Cedex, France \\
   \and Academia Sinica, Institute of Astronomy and Astrophysics, Taipei, Taiwan}
   \date{Received 2017 November 11; accepted 2017 December 29}

\abstract{ Observations by the Atacama Large Millimeter/sub-millimeter Array of the dust continuum and $^{13}$CO(3-2) millimeter emissions of 
the triple stellar system GG Tau A are analysed, giving evidence for a rotating gas disc and a concentric and coplanar dust ring. 
The present work complements an earlier analysis (Tang et al.~\cite{tang16}) by exploring detailed properties of the gas disc. 
A 95\% confidence level upper limit of 0.24 arcsec (34 au) is placed on the disc scale height at a distance of 1 arcsec (140 au) from the central stars. 
Evidence for Keplerian rotation of the gas disc is presented, the rotation velocity reaching $\sim$3.1 km\,s$^{-1}$ at 1 arcsec from the central stars, 
and a 99\% confidence level upper limit of 9\% is placed on a possible in-fall velocity relative contribution. 
Variations of the intensity across the disc area are studied in detail and confirm the presence of a hot spot in the south-eastern quadrant. However several 
other significant intensity variations, in particular a depression in the northern direction, are also revealed. 
Variations of the intensity are found to be positively correlated to variations of the line width. 
Possible contributions to the measured line width are reviewed, 
suggesting an increase of the disc temperature and opacity with decreasing distance from the stars.
\keywords{protoplanetary disks, stars: low-mass, stars: individual (GG Tau A)}
}
   \authorrunning{N.T. Phuong, P.N. Diep \& A. Dutrey et al., }  %author_head in even pages
   \titlerunning{Morphology of the $^{13}$CO(3-2) emission across the disc surrounding protostar GG Tau A}  % title_head in odd pages
   \maketitle
%% The author head (on even pages) and the title head (on odd pages) will be
%% automatically extracted from \author{} and \title{}. Whenever the title is too long,
%% you will be asked to supply a shorter one by inserting either \authorrunning{} or
%% \titlerunning{} before \maketitle. Anyway, you can specify your own heads.
%%
%% Note: In the following text body of your manuscript, please note several differences from
%%       other major journals:
%% (1) \subsection{Please Capitalize the First Letter of Each Notional Word in Subsection Title}
%% (2) Please Capitalize the First Letter of Each Notional Word in all tables' captions

%
%________________________________________________ sections below
%
\section{Introduction}           %% first-level sections will be auto-capitalized
\label{sect:intro}
GG Tau A is a triple stellar system, 1 to 5 million years old, located at 140 pc in a hole of the Taurus-Auriga star forming region. 
The separation between the main star GG Tau Aa and the close binary GG Tau Ab \mbox{(Ab1-Ab2)} is 35 au 
while the separation between GG Tau Ab1 and Ab2 is only 4.5 au \mbox{(Di Folco et al.~\cite{folco14}).} 
GG Tau A is surrounded by an envelope of gas and dust with a ring extending from $\sim$180 to 260 au 
and an outer disc extending up to $\sim$800 au from the central stars with an estimated mass of $\sim$0.15 solar masses 
(Dutrey et al.~\cite{dutrey94}). There is neither known molecular outflow nor jets associated with \mbox{GG Tau A}. 
Additional information about the system can be found in the review by Dutrey et al.~(\cite{dutrey16}) and from references therein. 
In particular high resolution Atacama Large Millimeter/sub-millimeter Array (ALMA) 
observations of $^{12}$CO(6-5) emission and underlying continuum (Dutrey et al.~\cite{dutrey14}) 
have suggested possible planet formation. The present article uses ALMA data of the $^{13}$CO(3-2) 
emission and underlying continuum that have been presented earlier by Tang et al.~(\cite{tang16}) 
together with $^{12}$CO(3-2) observations. Contrary to $^{12}$CO(3-2) emission, which extends down to 
small distances from the central stars, $^{13}$CO(3-2) emission is limited to an outer ring having an 
inner edge radius of $\sim$1 arcsec. The present analysis aims at complementing that of 
Tang et al.~(\cite{tang16}) by providing new detailed information on the properties of the gas disc.
% %
\section{OBSERVATIONS AND DATA REDUCTION}
\label{sect:Obs}

The observations used in the present article were made in cycle 1 of ALMA operation (2012.1.00129.S) on November 18 and 19, 2013 
in three blocks (Tang et al.~\cite{tang16}). The time spent on source was 1.84 hours. 
The number of antennas was 28, the longest baseline being 1284.3 m. 
The three blocks of continuum data have been merged and 
calibrated by the ALMA staff and the $^{13}$CO(3-2) data 
have been calibrated using CASA\footnote{http://casa.nrao.edu} and GILDAS\footnote{https://www.iram.fr/IRAMFR/GILDAS}. 
The origin of coordinates at RA=4h\,32m\,30.3s and DEC=$17^{\circ}\,31'\,40"$ corresponds to year 2000. 
Between 2000 and the time of observation, the source has moved by 0.24 arcsec east and \mbox{0.26 arcsec} south 
(proper motion of [17, $-$19] mas per year taken from SIMBAD\footnote{http://simbad.harvard.edu/simbad/sim-fbasic} database); 
the data have been corrected accordingly.

The continuum emission was observed at $\sim$0.9 mm wavelength 
over frequencies covering from 330.655 to 344.426 GHz. 
The beam size (FWHM) is $0.39\times0.29$ arcsec$^2$ with a position angle 
(measured from north to east) of $56^{\circ}$. The $^{13}$CO(3-2) emission was self-calibrated. 
The beam size (FWHM) is \mbox{$0.37\times0.31$ arcsec$^2$} with a position angle of 80$^\circ$.
The present analysis is performed in the image plane and we evaluate the uncertainty on position measurements
due to noise to be typically below 0.01 arcsec, depending on the signal to noise ratio. However, it is often dominated by systematics and needs to be evaluated for each case separately.    
The spectral resolution (channel) has been smoothed to 0.11 km\,s$^{-1}$ and the 
Doppler velocity covers between $-2$ and 15 km\,s$^{-1}$. Here, Doppler velocities are defined as the difference between 
the measured velocity and a systemic velocity of \mbox{$6.38\pm0.02$ km\,s$^{-1}$} as used by Dutrey et al. (\cite{dutrey14}) 
for $^{12}$CO(6-5) about which the profile is well symmetric.
 
\section{GENERAL FEATURES}
\label{sect:general}
\subsection{Continuum data}
Figure~1 (left) maps the brightness of the continuum emission. 
It shows an elliptical ring surrounding a central source. 
The right panels show the projections on the $x$ (right ascension offset) 
and $y$ (declination offset) axes of the central source intensity integrated over 
$y$ and $x$ respectively. Gaussian fits give mean values of 
0.06 and $-$0.13 arcsec and FWHM values of 0.40 and 0.33 arcsec in $x$ and $y$ respectively, 
similar to the beam size: the central source is unresolved.
% ------------------------------------------------------------
%      A figure as large as the width of the column
%-------------------------------------------------------------
%=======FIGURE 1 ==========
   \begin{figure}[!htbp]
   \centering
   \includegraphics[width=4.1cm,trim=2cm 0.5cm 1cm 0.5cm]{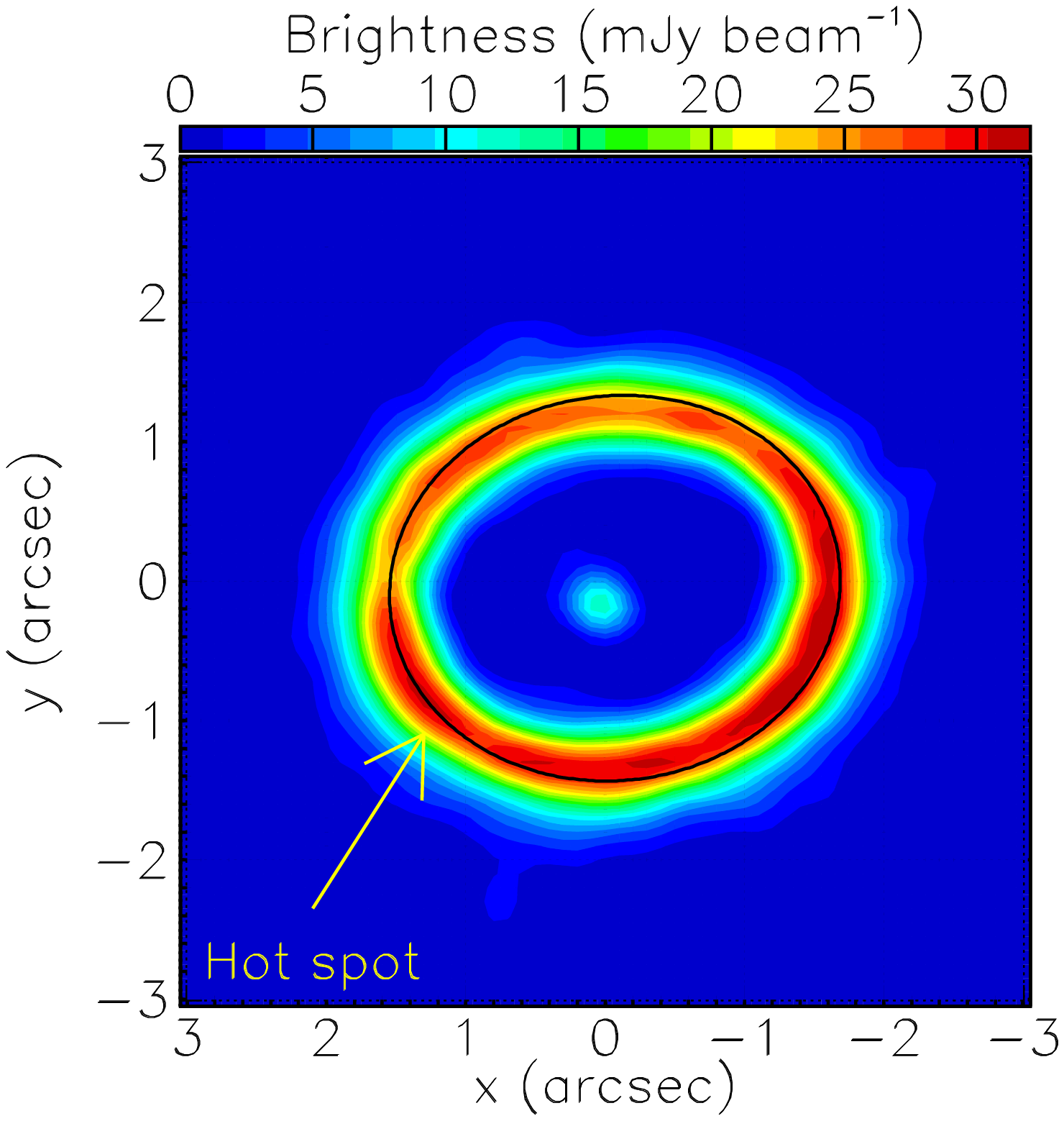}
   \includegraphics[width=5.3cm,trim=0cm 2.1cm 0cm 0.5cm]{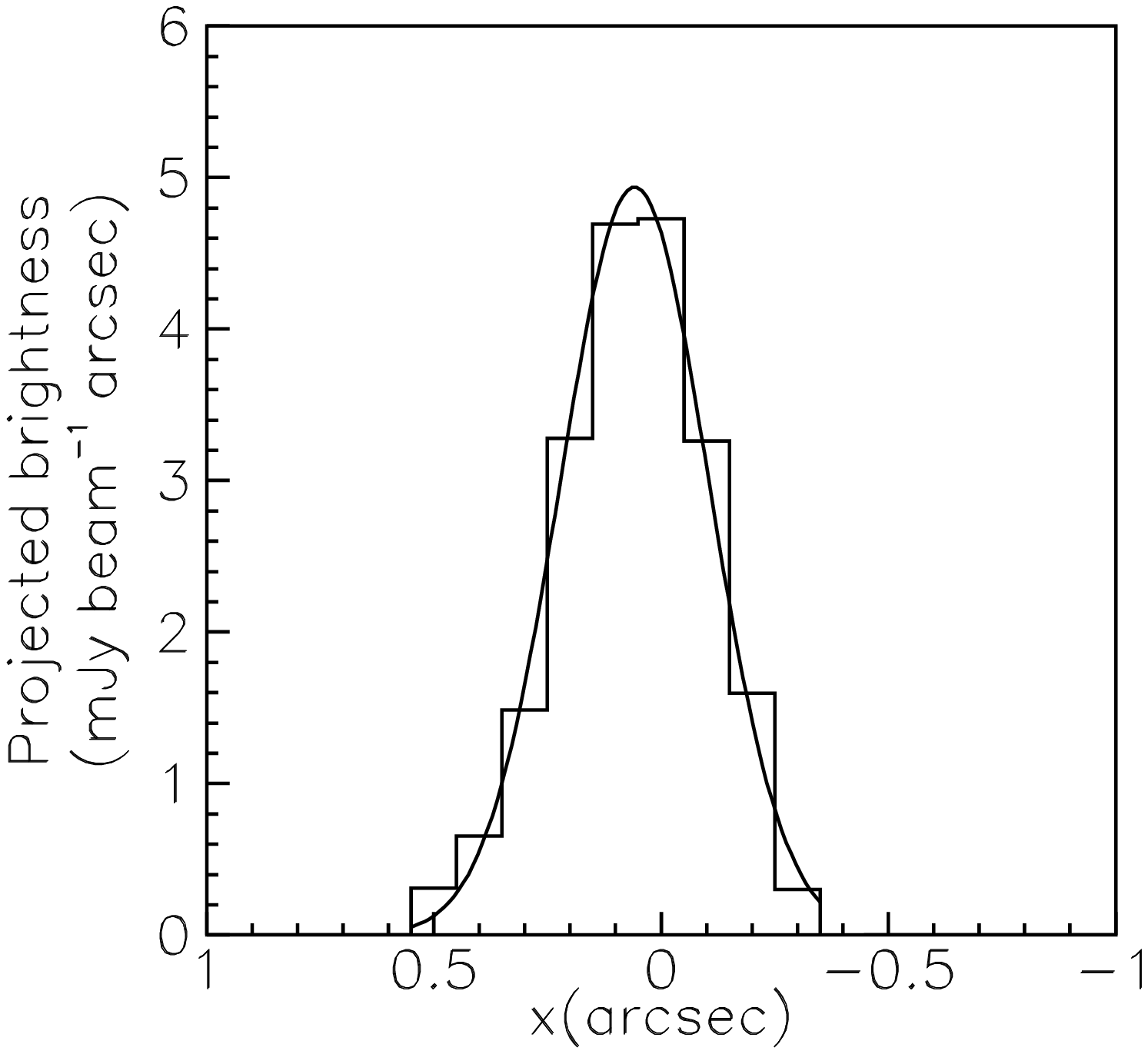}
   \includegraphics[width=4.45cm,trim=2.5cm 2.1cm 0cm 0.5cm]{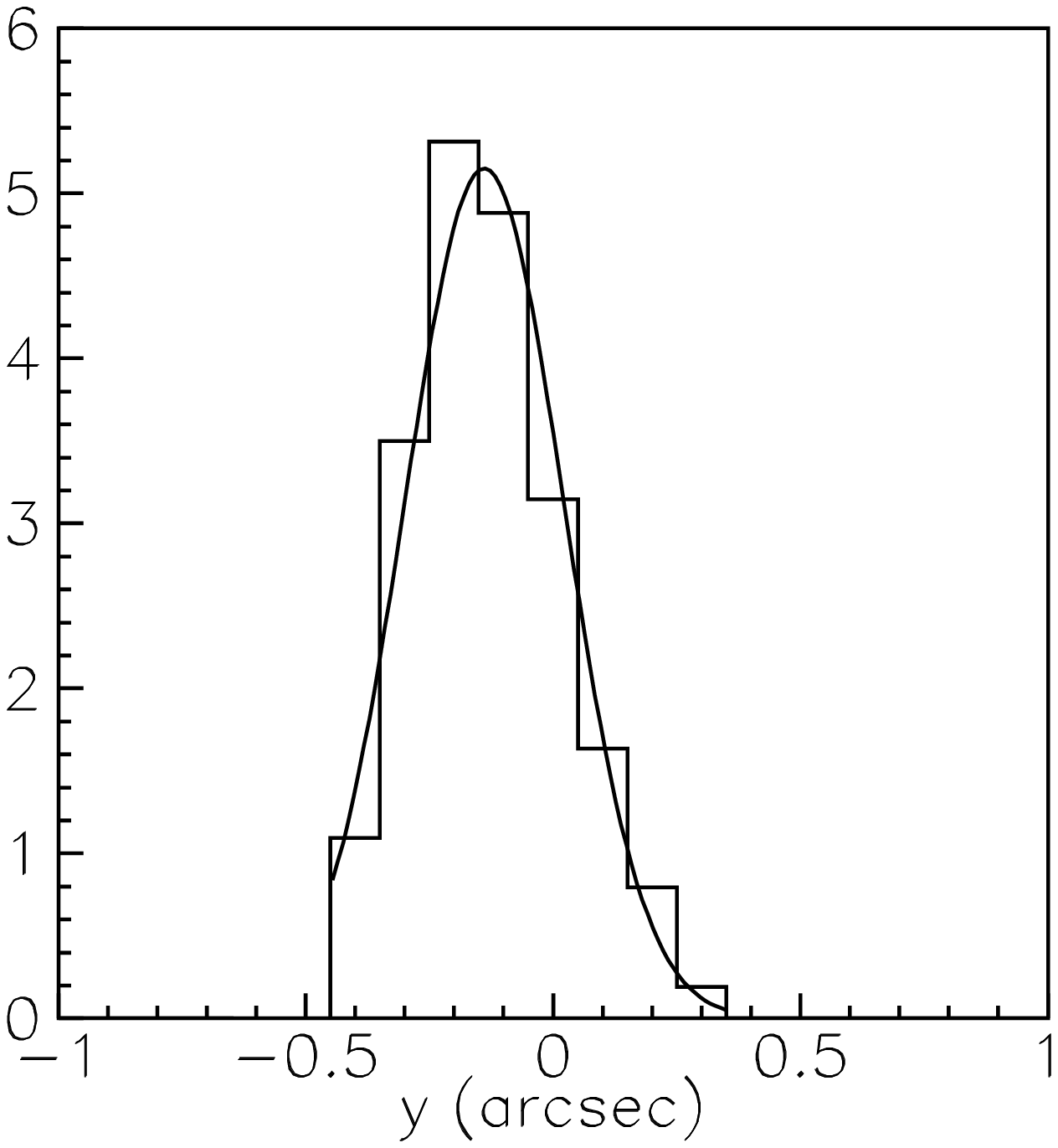}
   \caption{Brightness of the dust ring continuum emission. {\it{Left}}: sky map; 
the black ellipse is the fit to $<R>$ shown in Figure 3; 
the yellow arrow points to the region of the hot spot observed by Dutrey et al. (\cite{dutrey14}) 
and Tang et al. (\cite{tang16}) in $^{12}$CO(6-5) and $^{12}$CO(3-2) emissions. 
{\it{Middle and right}}: projections on the $x$ and $y$ axes of the central 
source brightness integrated over $y$ and $x$ respectively. The lines show Gaussian best fits. }
   \label{Fig1}
   \end{figure}
%=======FIGURE 2 ==========
\begin{figure}[!htbp]
   \centering
   \includegraphics[width=4.4cm,trim=1.5cm 1.75cm 1cm 1.75cm]{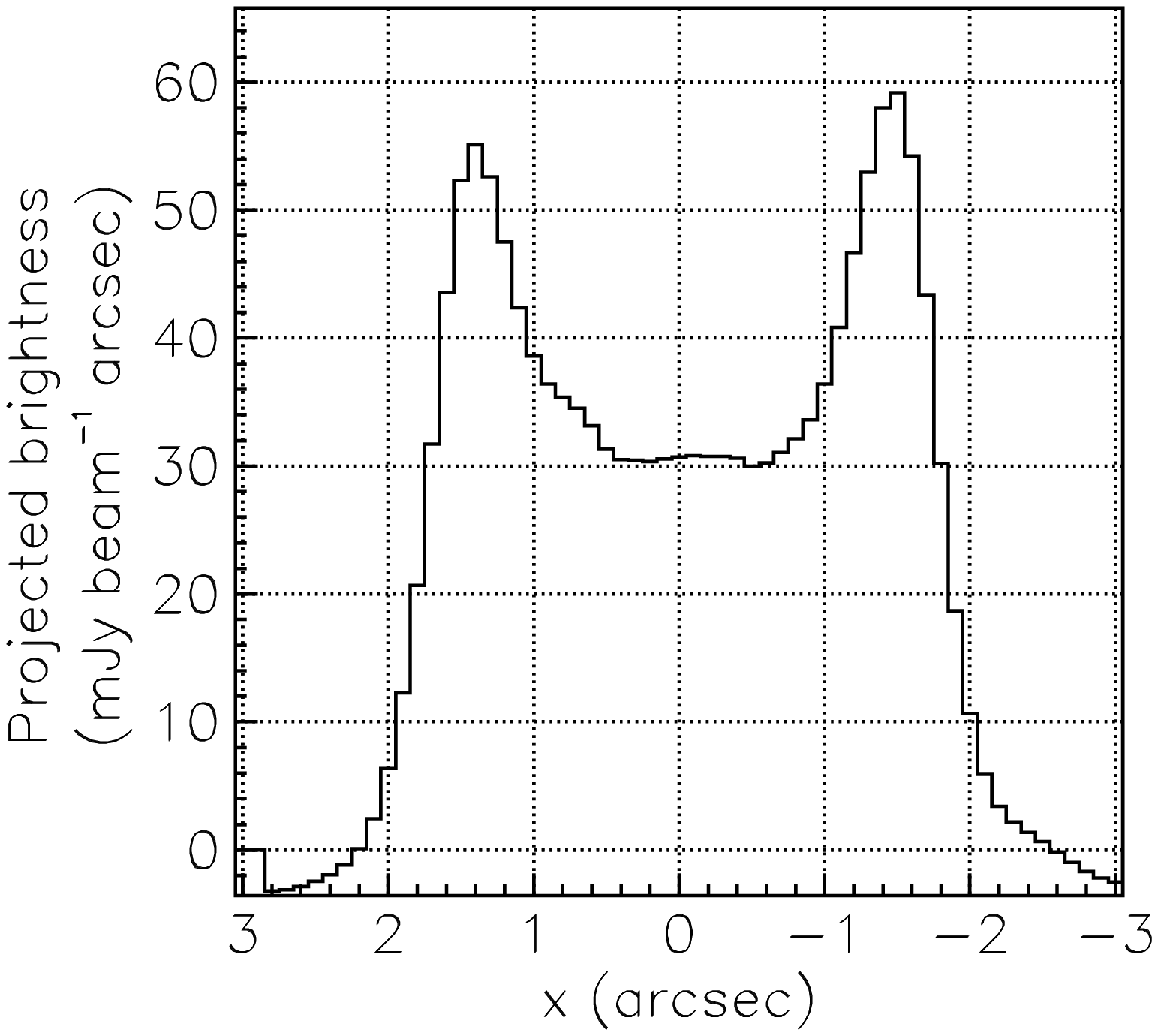}
   \includegraphics[width=4.4cm,trim=1.5cm 1.75cm 1cm 1.75cm]{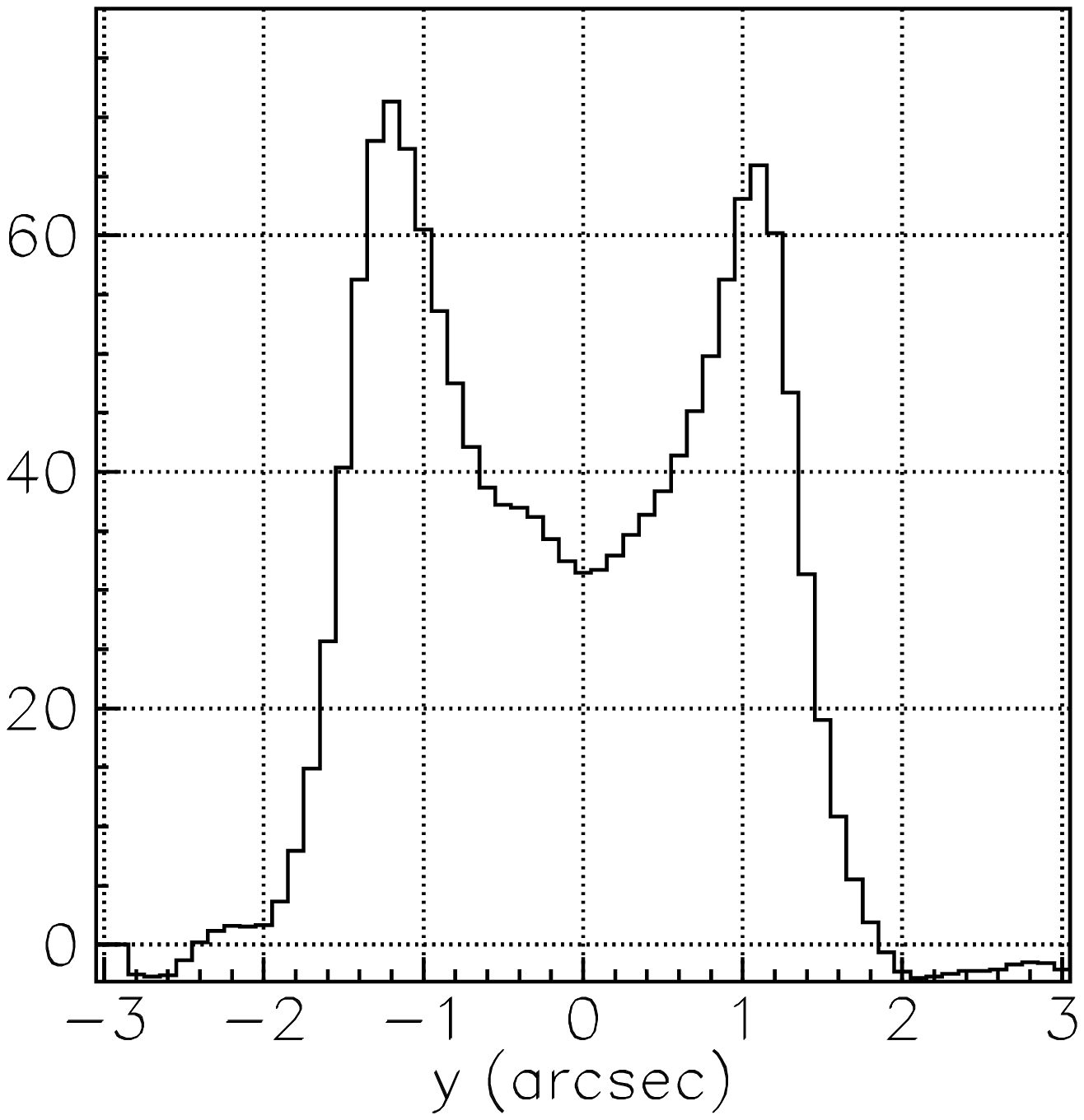}
   \includegraphics[width=4.8cm,trim=.25cm 1.75cm 1cm 1.75cm]{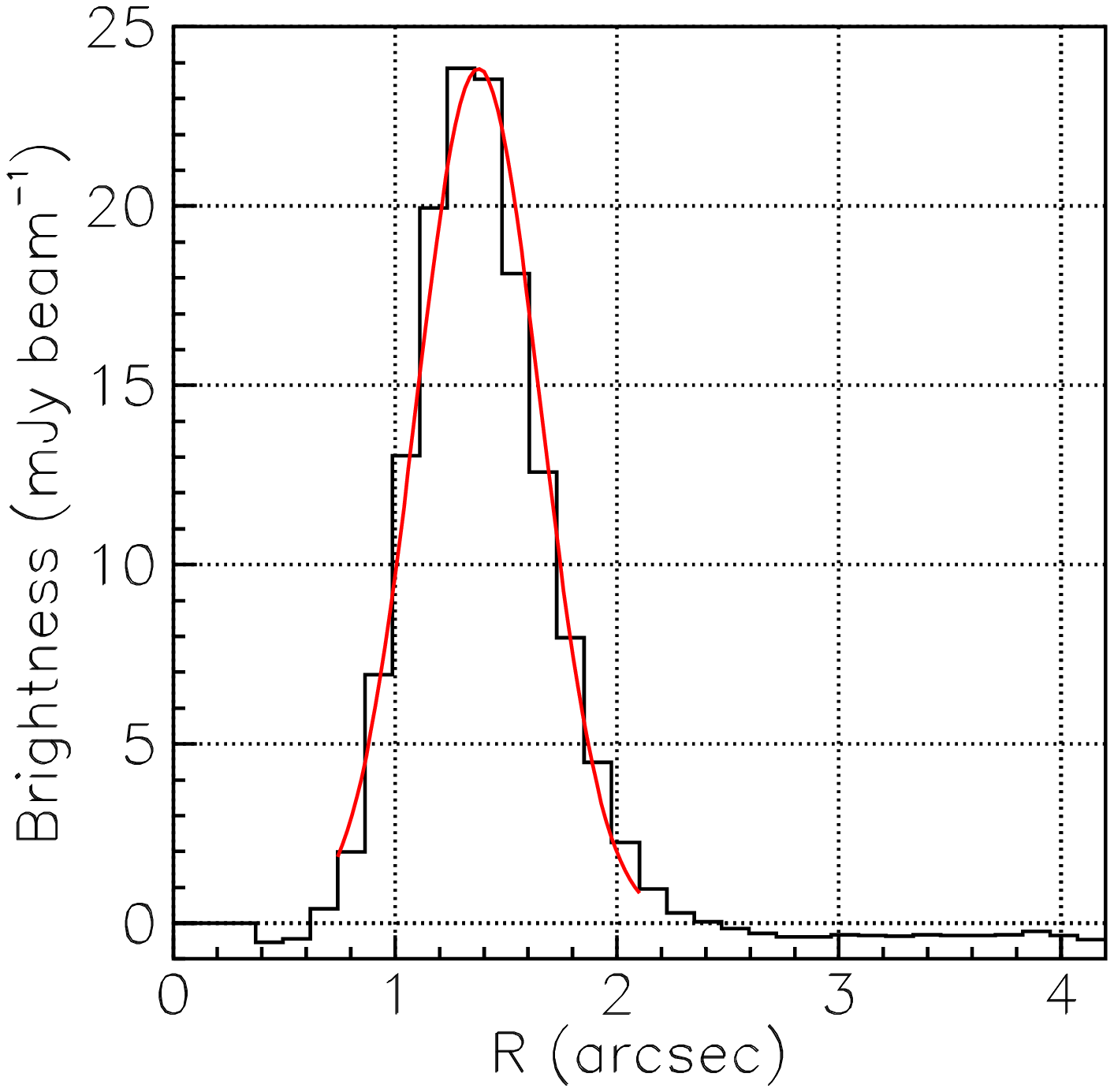}
   \caption{Continuum brightness of the dust ring emission projected 
on the $x$ (left) and $y$ (middle) axes and integrated over $y$ and $x$ respectively. 
The right panel shows its distribution as a function of $R$, 
averaged over $\varphi$, together with the Gaussian best fit to the peak. 
In all three panels pixels having $R'<0.5$ arcsec are excluded.}
   \label{Fig2}
   \end{figure}

Figure~2 (left) displays the $x$ and $y$ projections of the 
continuum brightness integrated over $y$ and $x$ respectively. 
It requires the distance $R'=\sqrt{(x-0.06)^2+(y+0.13)^2}$ to the central source
to exceed 0.5 arcsec, thereby excluding its contribution. 
The corresponding mean values of $x$ and $y$ are $-0.05$ and $-0.09$ arcsec respectively, 
showing that the ring is shifted north-west by $\sim$0.12 arcsec with respect to the central source. 
The position and width measurements illustrated in Figure~1 and Figure~2 are accurate to be better than 0.02 arcsec, using the residual of the fits to estimate 
measurement errors. They are dominated by systematics rather than simply by thermal noise. However, the angular separation between GG Tau Aa and Ab is 0.25 arcsec: depending on 
what is being talked about, the position of the "centre" may vary by some $\pm$0.1 arcsec.  
Figure~2 (right) displays the dependence on $R=\sqrt{x^2+y^2}$ of the continuum brightness 
averaged over position angle \mbox{$\varphi=90^{\circ}-tan^{-1}(y/x)$} (measured from north to east), 
again excluding the central source by requiring $R'>0.5$ arcsec. 
A Gaussian fit to the peak gives a mean of $1.445\pm0.015$ arcsec and a $\sigma$ of $0.266\pm0.015$ arcsec depending on the interval of $R$ over which the fit is performed. 
Retaining a $\sigma$ value of 0.266 arcsec and subtracting the beam size in quadrature gives a de-convolved FWHM of $0.528\pm0.035$ arcsec. 
Tang et al.~(\cite{tang16}) quote a value of 0.54 arcsec for the de-projected and de-convolved width of a uniform ring. 
The effect of de-projection is negligible and the correcting factor for Gaussian to square box fit is $\sqrt{2\pi}/{2\sqrt{2 \ln2}}=1.06$, bringing the Tang et al.~(\cite{tang16}) value down to 0.51 arcsec to be compared with the present result of $0.528\pm0.035$ arcsec. This is a very good agreement given that the ring is neither uniform nor perfectly Gaussian and that possible wings of faint emission beyond 1.8 arcsec would affect differently the two fitting procedures.

Figure~3 (left) displays the mean value of $R$, $<R>$, weighted by the radial average of the brightness across the ring over the interval $1<R<2$ arcsec.
A fit of the dependence of $<R>$ on $\varphi$ as an ellipse of semi-major and semi-minor axes $a_0$ and $b_0$ and offset by $\Delta x$ and $\Delta y$
has been made to first order in the offsets and in the ellipticity:

\begin{equation}
r=(\frac{\cos^2(\varphi-\varphi_0)}{a_0^2}-2 \frac{\Delta x}{\sqrt{a_0b_0}}\frac{\cos(\varphi-\varphi_0)}{a_0^2}+\frac{\sin^2(\varphi-\varphi_0)}{b_0^2}
-2\frac{\Delta y}{\sqrt{a_0b_0}}\frac{\sin(\varphi-\varphi_0)}{b_0^2})^{-1/2}
\end{equation}

It gives $a_0=1.62$ arcsec and $b_0=1.38$ arcsec, position angle of the major axis $\varphi_0=97^{\circ}$ 
and small offsets $\Delta x=-0.07$ and $\Delta y=-0.05$ arcsec, at the level of measurement uncertainties. 
This confirms the good centering of the ring on the origin of coordinates and the aspect ratio corresponds 
to a tilt with respect to the sky plane $\theta=\cos^{-1}(1.38/1.62)=32^{\circ} \pm 4^\circ$ of a circular 
ring about the rotated (by $7.0^{\circ}$) $x$ axis.
%
%=======FIGURE 3 ========== 
\begin{figure}[!htbp]
   \centering
   \includegraphics[width=4.5cm,trim=.5cm 2.3cm 0cm 0.5cm]{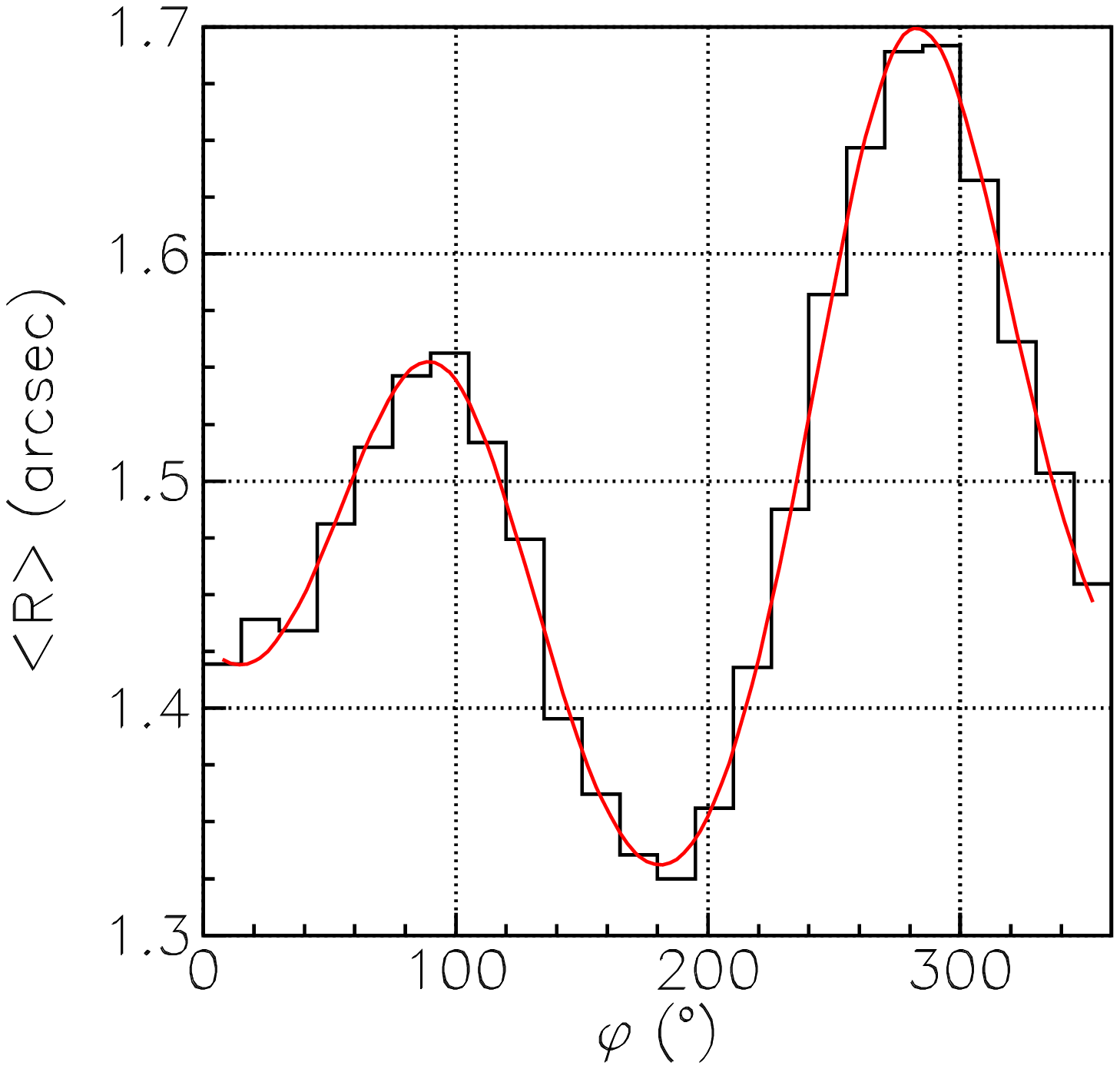}
   \includegraphics[width=4.5cm,trim=.5cm 2.cm 0cm 0.5cm]{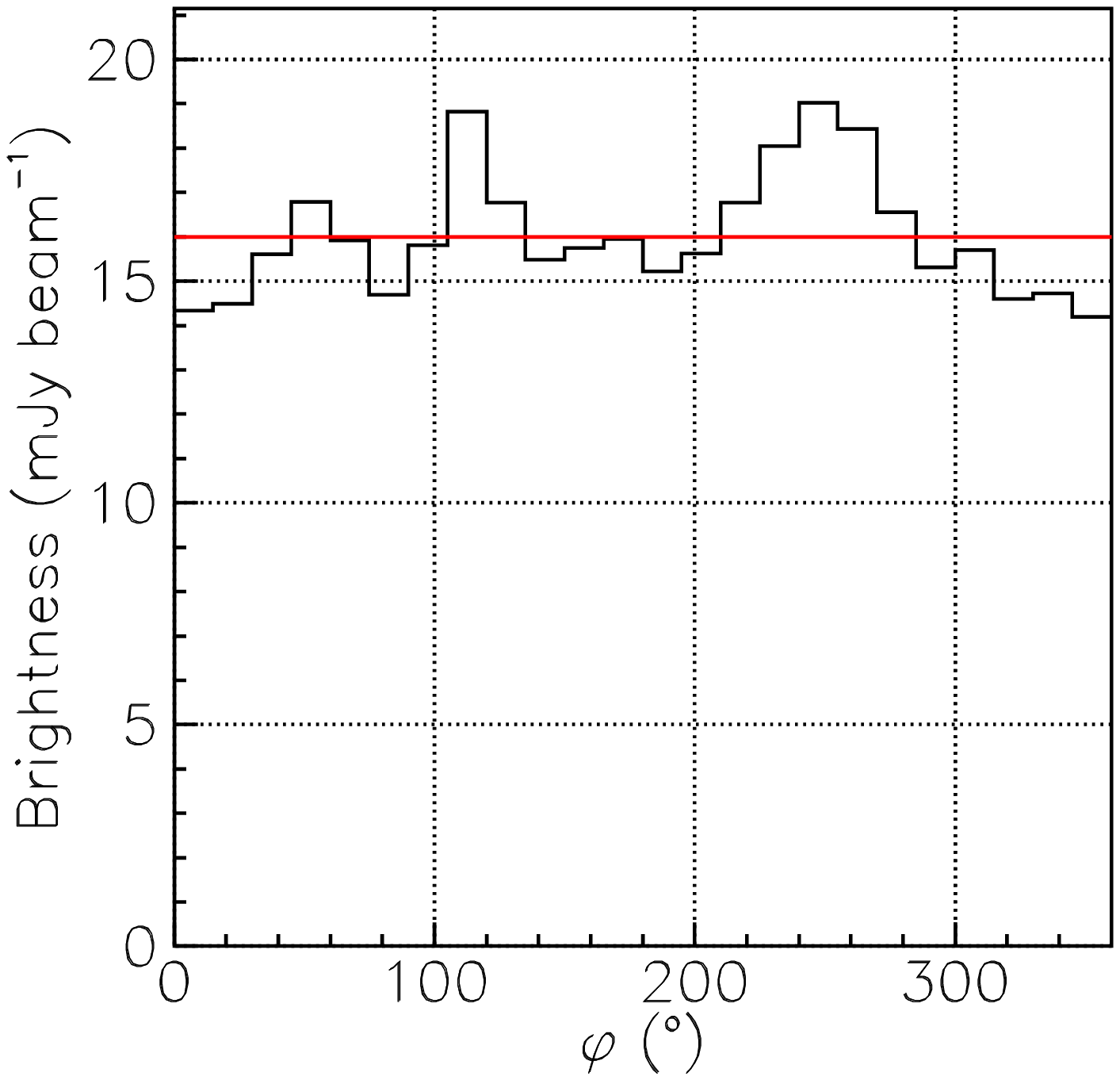}
   \caption{Continuum emission. {\it{Left:}} Dependence on $\varphi$ of $<R>$ 
calculated in the interval $1<R<2$ arcsec. The red line is the best fit to an elliptical tilted ring offset from the origin (see text). 
{\it{Right:}} Dependence on $\varphi$ of the disc plane continuum brightness averaged over $R$ 
in the interval $1<R<2$ arcsec. The red line shows the mean value. }
   \label{Fig3}
   \end{figure}

These results confirm the values quoted by Tang et al.~(\cite{tang16}): 1.63 arcsec instead of 1.62 arcsec for $a_0$, 
7.5$^{\circ}$ instead of 7.0$^{\circ}$ for the position angle and 36.4$^{\circ}$ instead of 32$^{\circ}$ 
for the tilt with respect to the sky plane, the latter being measured to no better than $\pm 4^\circ$. 
The values quoted for the tilt by Dutrey et al.~(\cite{dutrey14}) 
are $37^{\circ}\pm1^{\circ}$ for $^{12}$CO(6-5) and $35.0^{\circ}\pm0.2^{\circ}$ for the dust. 
Figure~3 (right) displays the dependence on position angle $\varphi$ of the 
continuum brightness averaged over $R$ in the interval $1<R<2$ arcsec. 
Here we have used the fact that the ratio between the beam area in the sky plane and its de-projected value in the disc plane is equal to $<R>/a_0$. 
In the disc plane the brightness is uniform over the disc circumference and equal to 16.0 mJy beam$^{-1}$ to within $\pm$8.5\% (rms).

\subsection{$^{13}$CO(3-2) line emission}
Figure~4 (left) displays the brightness distribution over the data cube. 
A Gaussian fit to the noise peak gives a mean of $-0.19$ mJy\,beam$^{-1}$ and a $\sigma$ of 7.2 mJy\,beam$^{-1}$ (0.56 K). 
Figure~4 (right) displays the Doppler velocity ($V_z$) spectrum integrated over $8\times8$ arcsec$^2$, 
with a double-horn profile typical of a rotating volume. It is centred to better than 0.1 km\,s$^{-1}$. 
In what follows, throughout the article, we restrict the Doppler velocity range to $|V_z|<2$ km\,s$^{-1}$ unless specified otherwise. 

%=======FIGURE 4 ==========
\begin{figure}
   \centering
   \includegraphics[width=5.cm,trim=1.5cm 1.5cm 1cm 0.5cm]{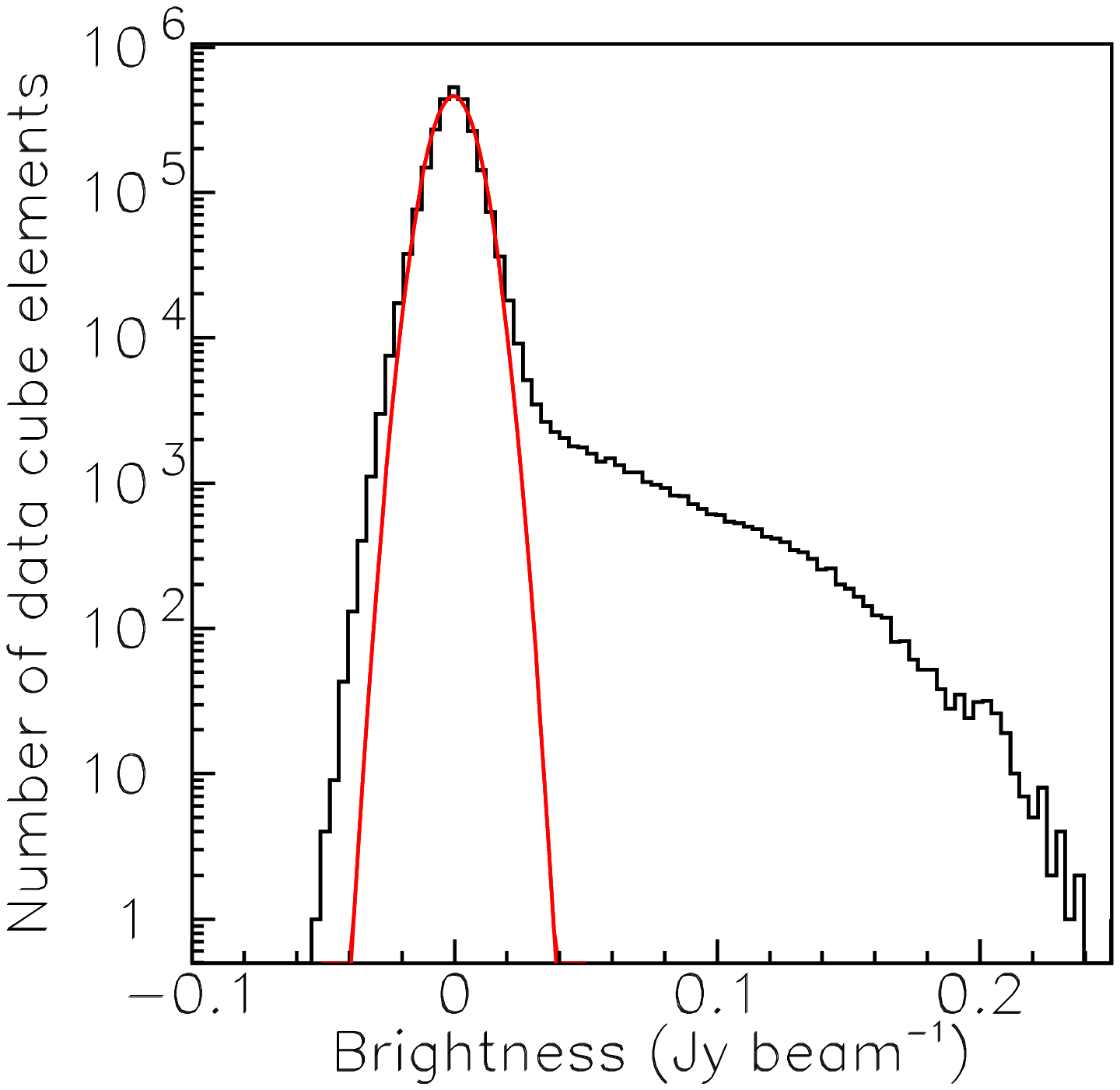}
   \includegraphics[width=5.cm,trim=0cm 1.5cm 2.5cm 0.5cm]{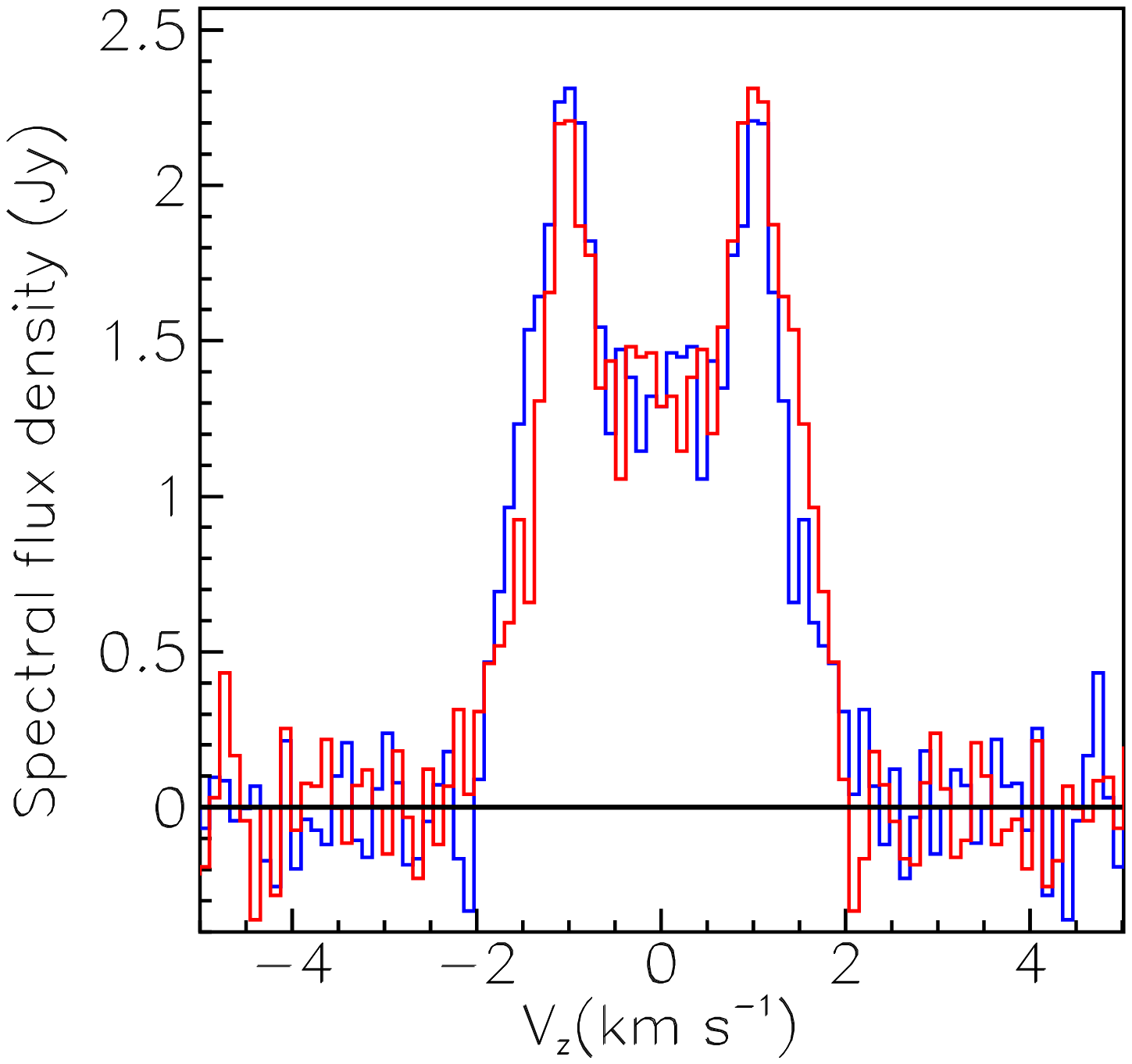}
   \caption{{\it{Left:}} Line brightness distribution (Jy\,beam$^{-1}$); the red curve is a Gaussian fit to the noise peak. 
{\it{Right:}} Doppler velocity spectrum weighted by brightness and integrated over $8\times8$ arcsec$^2$ (blue); 
the red histogram is obtained from the original by symmetry about the origin.}
   \label{Fig4}
   \end{figure}

Figure~5 displays the sky maps of the velocity-integrated brightness, or integrated intensity, and of the mean Doppler velocity. 
The map of the integrated intensity shows a clear ring of gas surrounding the central stars and having morphology similar to the dust morphology, 
indicating a concentric circular gas disc having the same inclination as the dust ring on the sky plane. It displays no central emission, 
with an abrupt inner cut-off at $\sim$1 arcsec; there is no significant emission inside an ellipse scaled down from the dust ellipse by a factor $\sim$3, 
meaning a de-projected radius of $\sim1.62/3=0.54$ arcsec. 
The velocity map excludes the region inside the scaled-down ellipse where noise dominates. 
It displays a clear velocity gradient along the major axis of the ellipse, as expected from rotation of the tilted disc about its axis. 
Note that an in-falling (rather than rotating) disc would display instead a gradient along the minor axis of the ellipse. 
In general adding some in-fall motion to rotation would cause the axis of the velocity gradient to deviate from the major axis, 
the more so the larger the relative contribution of in-fall.

%=======FIGURE 5 ==========
\begin{figure}
   \centering
   \includegraphics[width=5.cm,trim=1.cm 0.75cm 2cm 1.5cm]{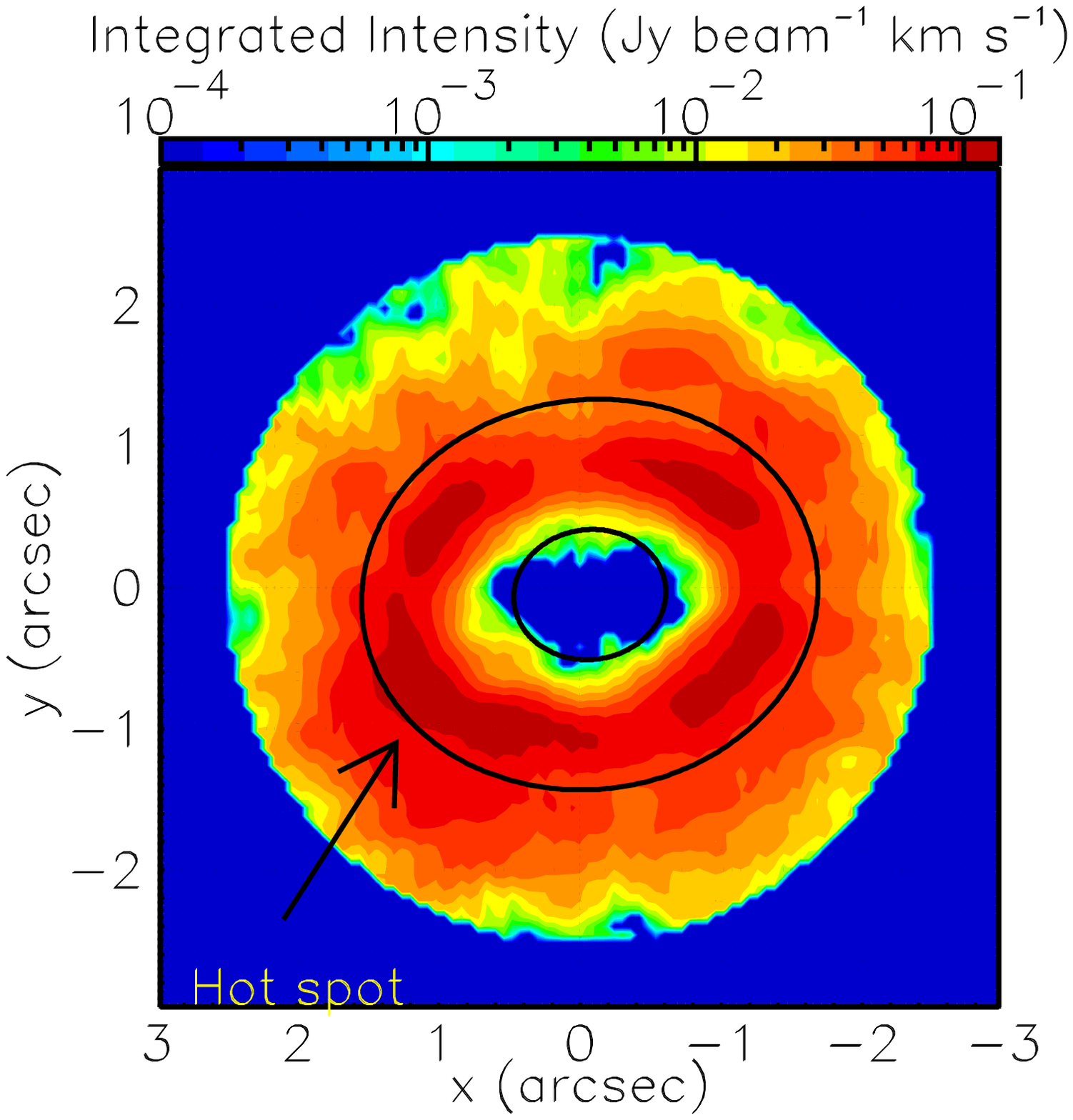}
   \includegraphics[width=5.65cm,trim=-.5cm 0.75cm 2cm 1.2cm]{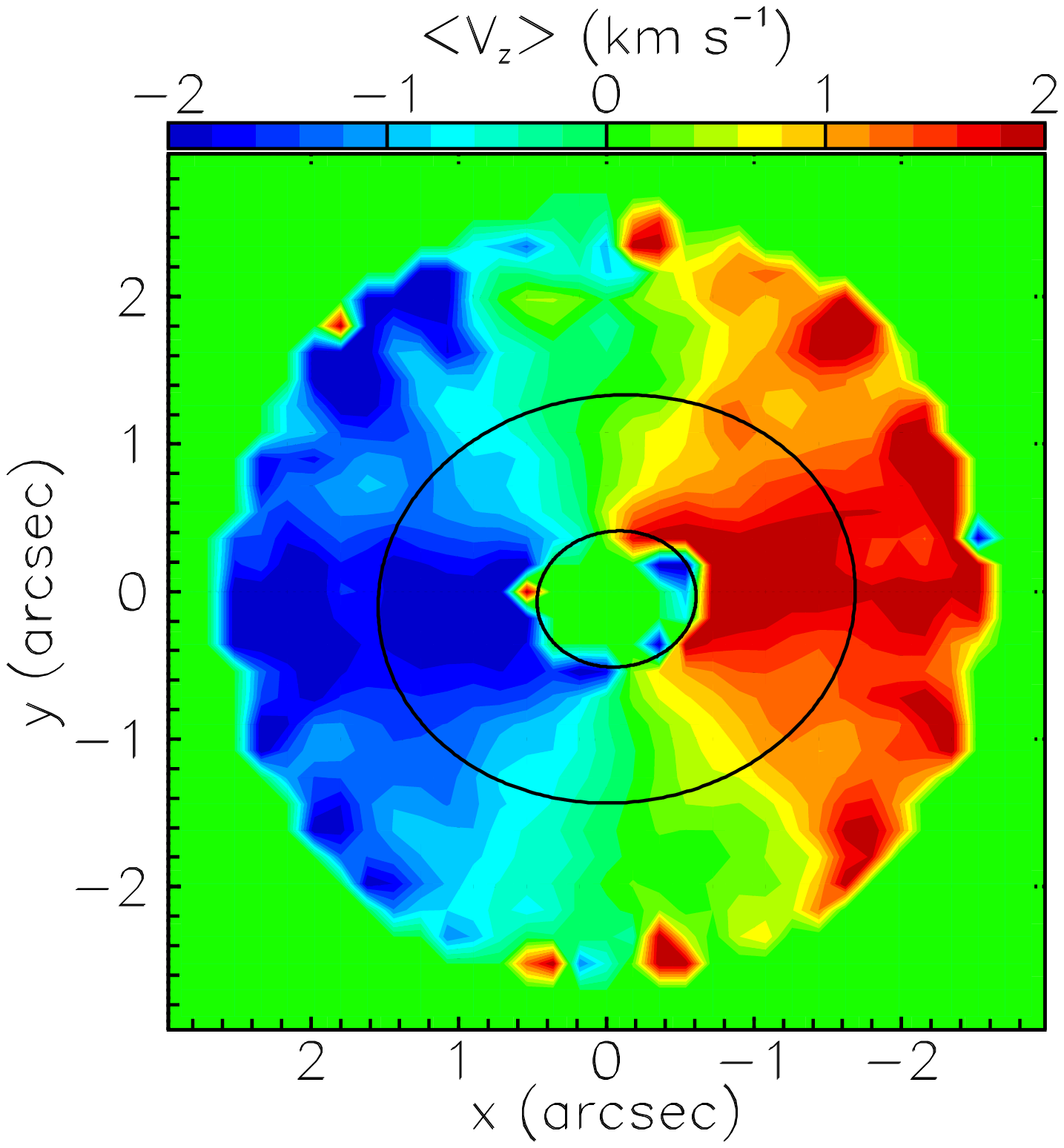}
   \caption{{\it{Left:}} Sky map of the $^{13}$CO(3-2) integrated intensity. 
The black arrow shows the position of the hot spot in $^{12}$CO(6-5) 
(Dutrey et al., \cite{dutrey14}) and $^{12}$CO(3-2) (Tang et al., \cite{tang16}). 
{\it{Right:}} Sky map of the mean Doppler velocity (weighted by brightness) 
excluding the region contained in the scaled-down ellipse shown in the left panel. 
In both panels $R<2.5$ arcsec and the black ellipses are the best fit 
to the distribution of $<R>$ in the continuum data and its 
scaled-down version (by a factor 3). }
   \label{Fig5}
   \end{figure}

Figure~6 is the equivalent for the line of Figure~2 for the continuum: 
projections on the $x$ and $y$ axes and $r$-dependence, averaged over $\varphi$, 
of the integrated intensity, where $r$ is now the de-projected value of $R$ 
in the disc plane (see Figure~8). Here, de-projection assumes a tilt angle of $32^\circ$ and 
a position angle of the disc axis of $7^\circ$, as for the dust. 
In all panels we exclude the central region where noise dominates by requiring $r>0.54$ arcsec. 
When compared with the dust (continuum) ring, the gas (line) ring is broader 
and peaks at smaller radii. The mean values of $x$ and $y$ are 0.02 and 
$-0.10$ arcsec respectively. A fit to the integrated intensity 
distribution as a function of $r$ as a sum of three Gaussians 
is shown in the right panel of the figure. 
The means and widths of the Gaussians are fixed to the values 
obtained by Tang et al. (\cite{tang16}) when fitting the western half of the gas disc. 

%=======FIGURE 6 ==========
\begin{figure}
   \centering
   \includegraphics[width=4.7cm,trim=0.cm 1.5cm 0.5cm 0cm]{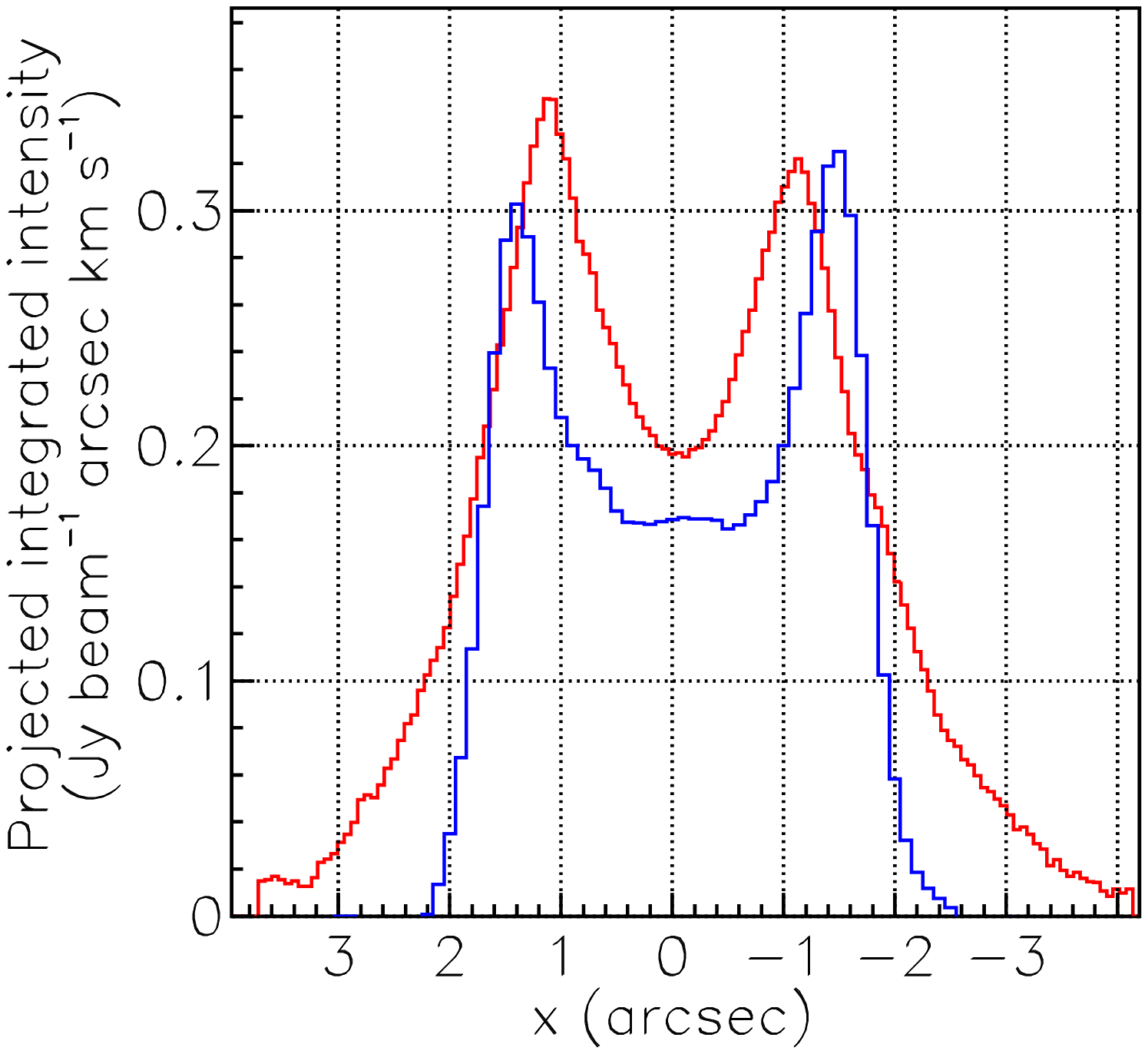}
   \includegraphics[width=4.13cm,trim=2.cm 1.5cm 0.5cm 0cm]{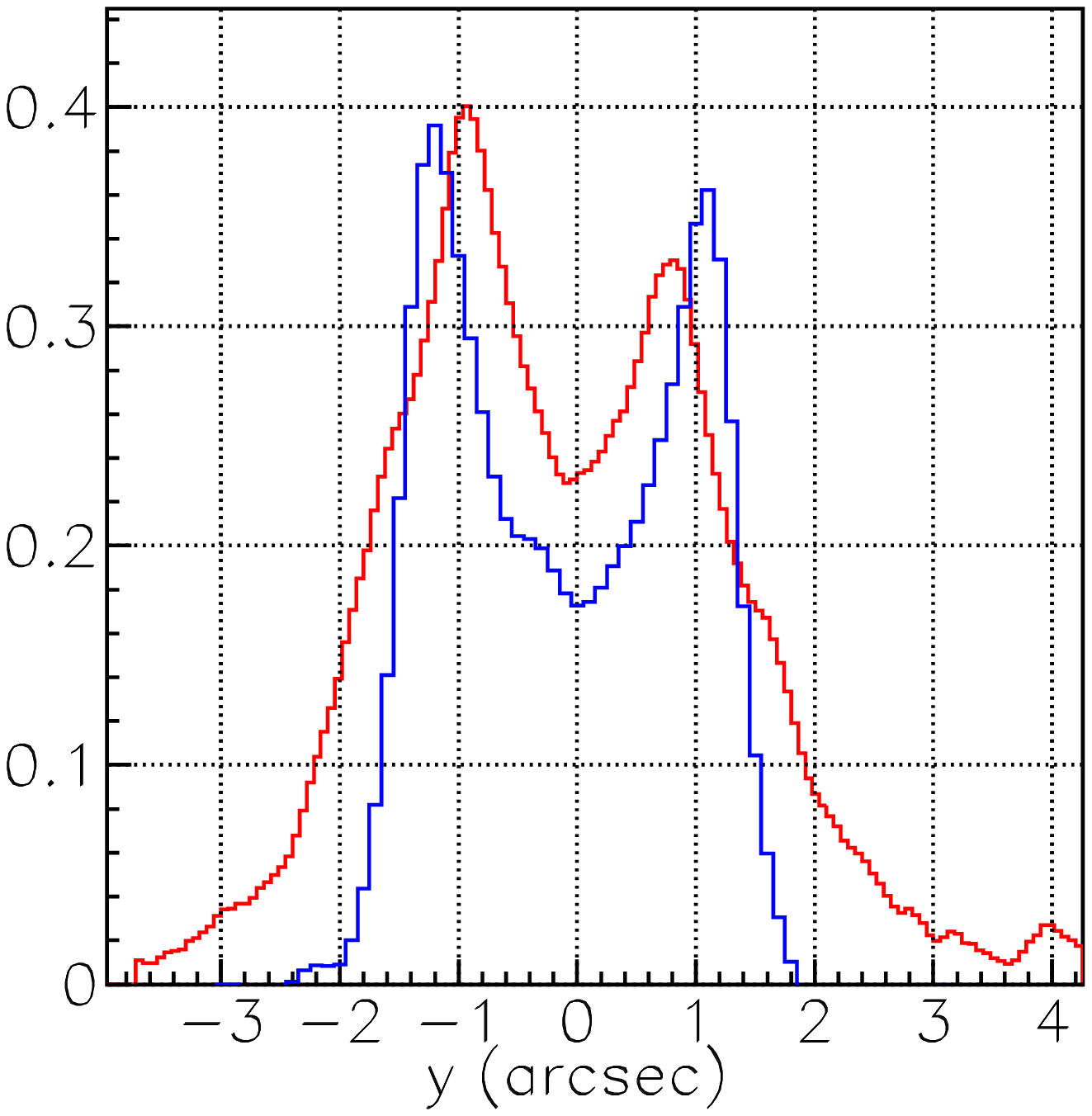}
   \includegraphics[width=4.7cm,trim=0.cm 1.5cm 0.5cm 0cm]{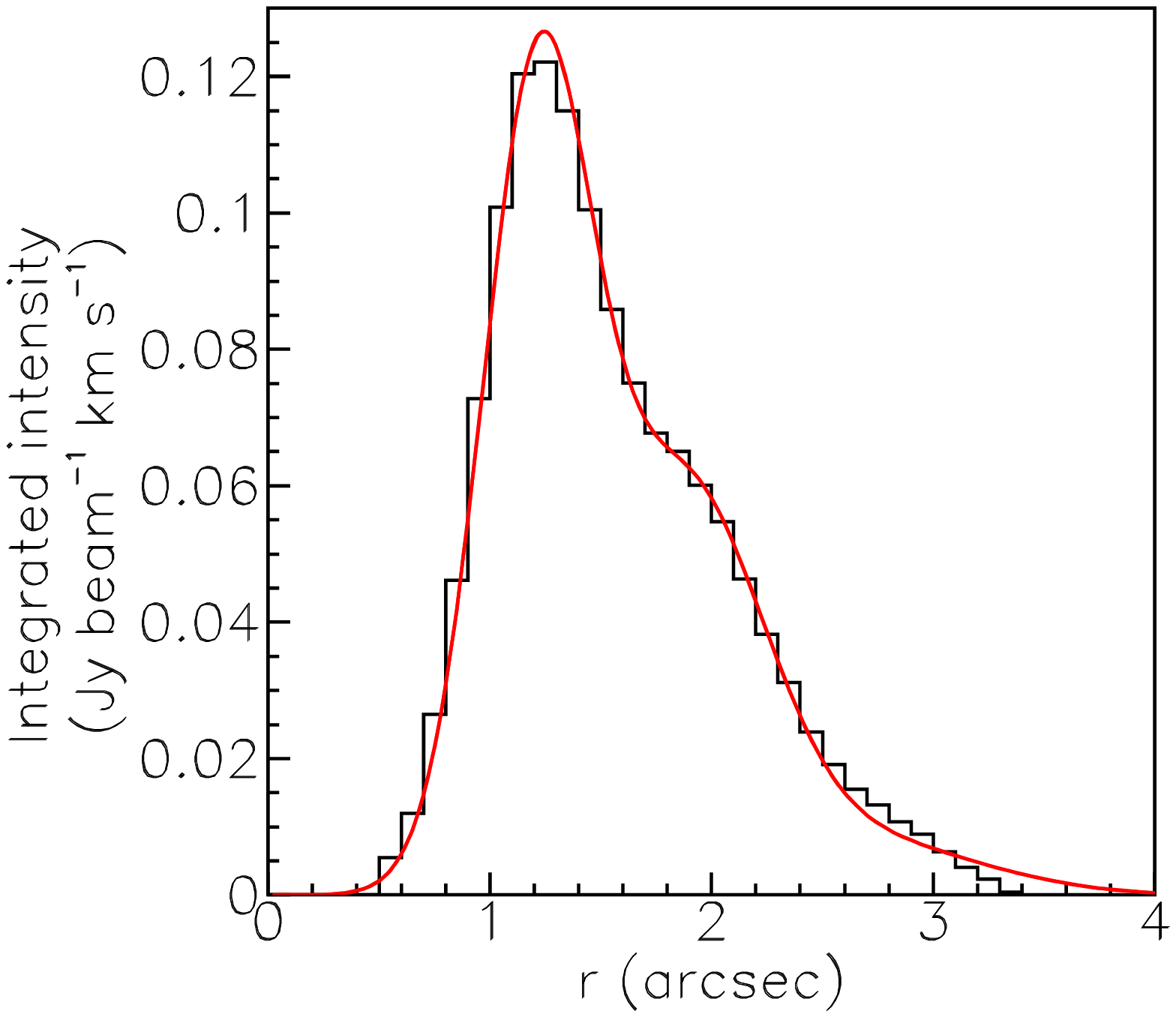}
   \caption{Line emission. {\it{Left}} and {\it{middle:}} Continuum brightness (blue, arbitrary normalisation) and line integrated intensity (red) projected on the $x$ (left) and $y$ (middle) axes in the region of $r>0.5$ arcsec. 
{\it{Right:}} $r$-dependence, averaged over $\varphi$, of the integrated intensity in the disc plane. The red line is a fit using the same three Gaussians as in Tang et al.~(\cite{tang16}).}
   \label{Fig6}
   \end{figure}
%=======FIGURE 7 ==========
\begin{figure}
   \centering
   \includegraphics[width=4.5cm,trim=1cm 1.75cm 0cm 0.5cm]{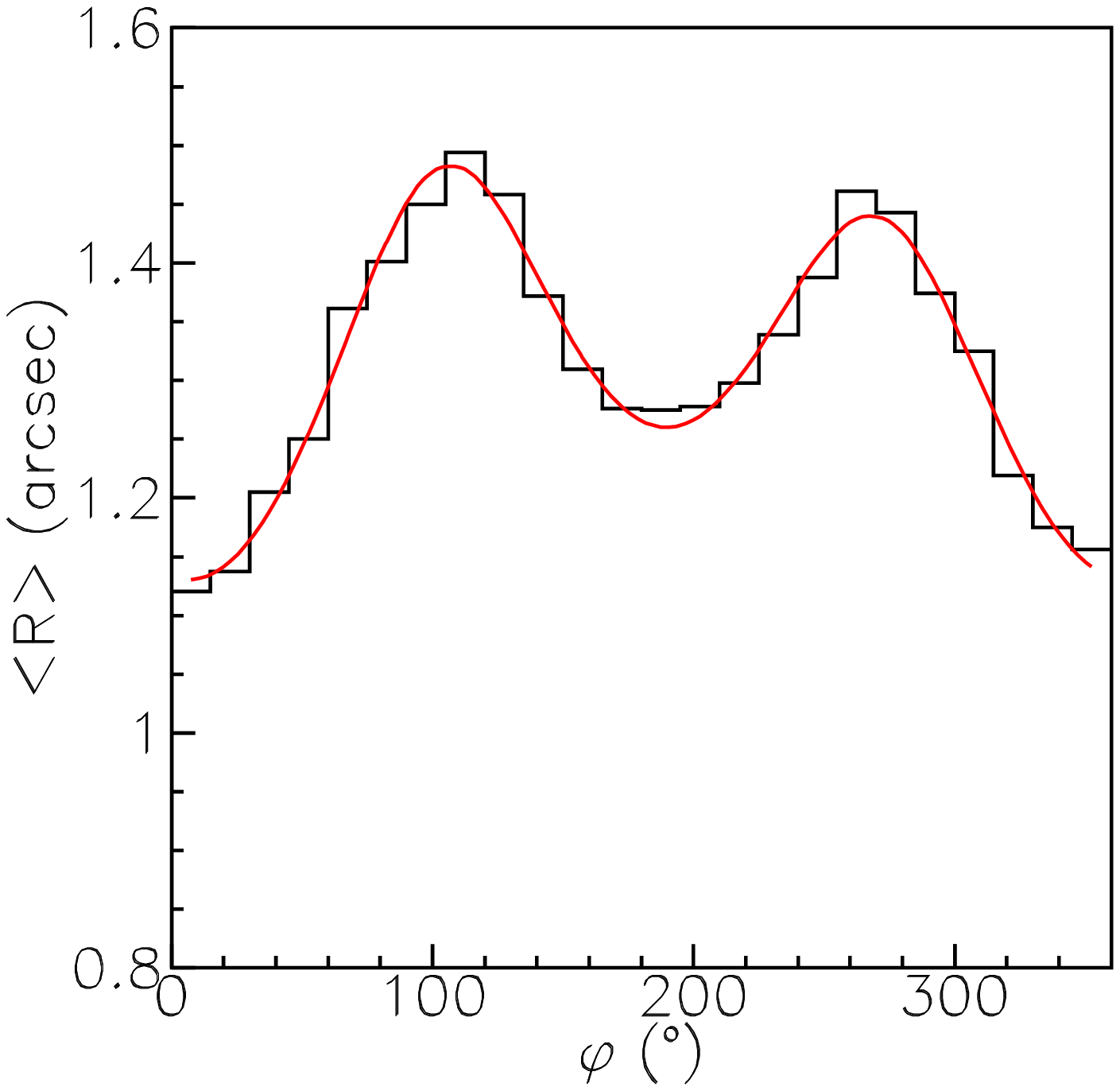}
   \includegraphics[width=4.5cm,trim=1cm 1.75cm 0cm 0.5cm]{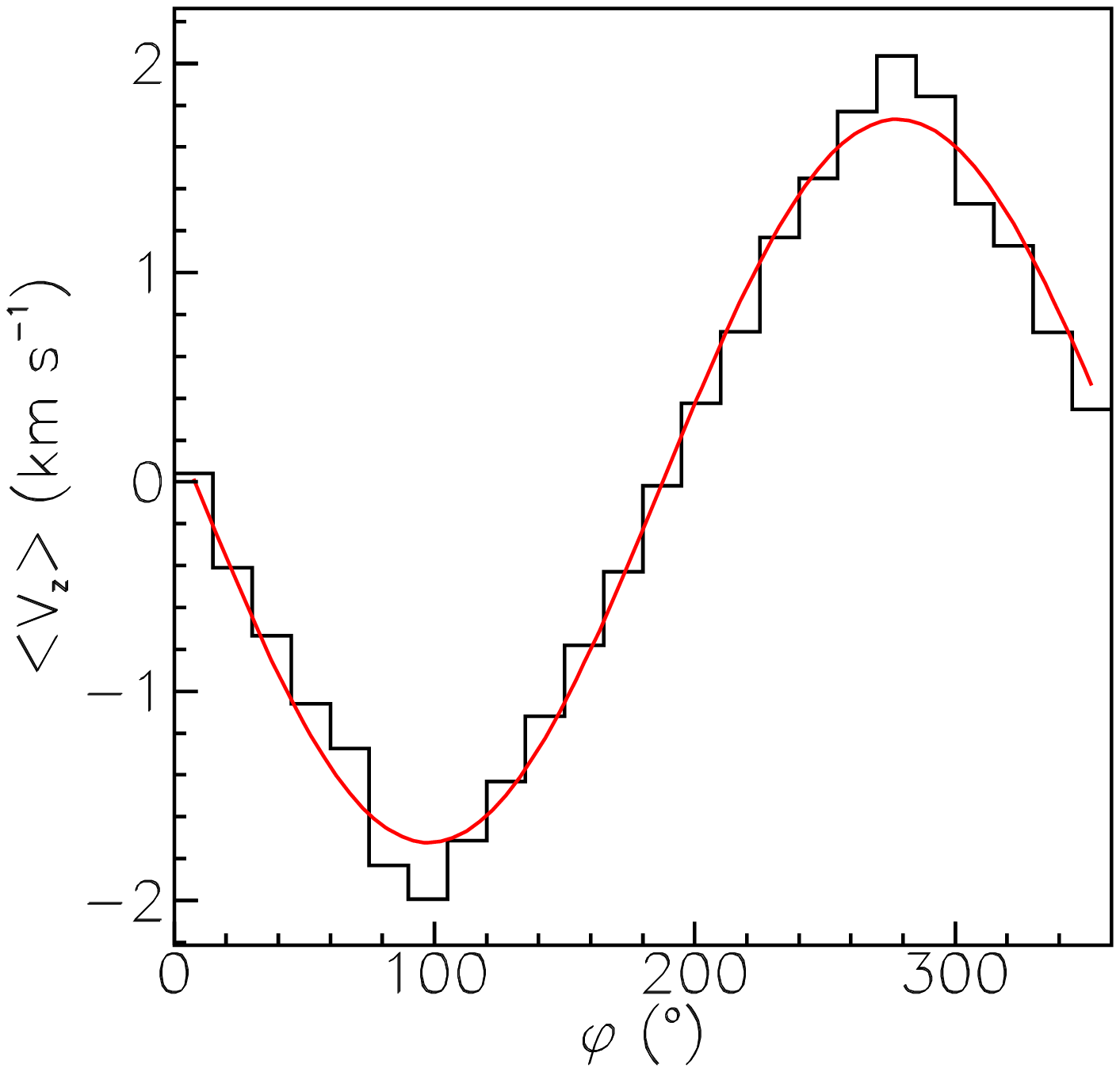}
   \includegraphics[width=4.5cm,trim=1cm 1.75cm 0cm 0.5cm]{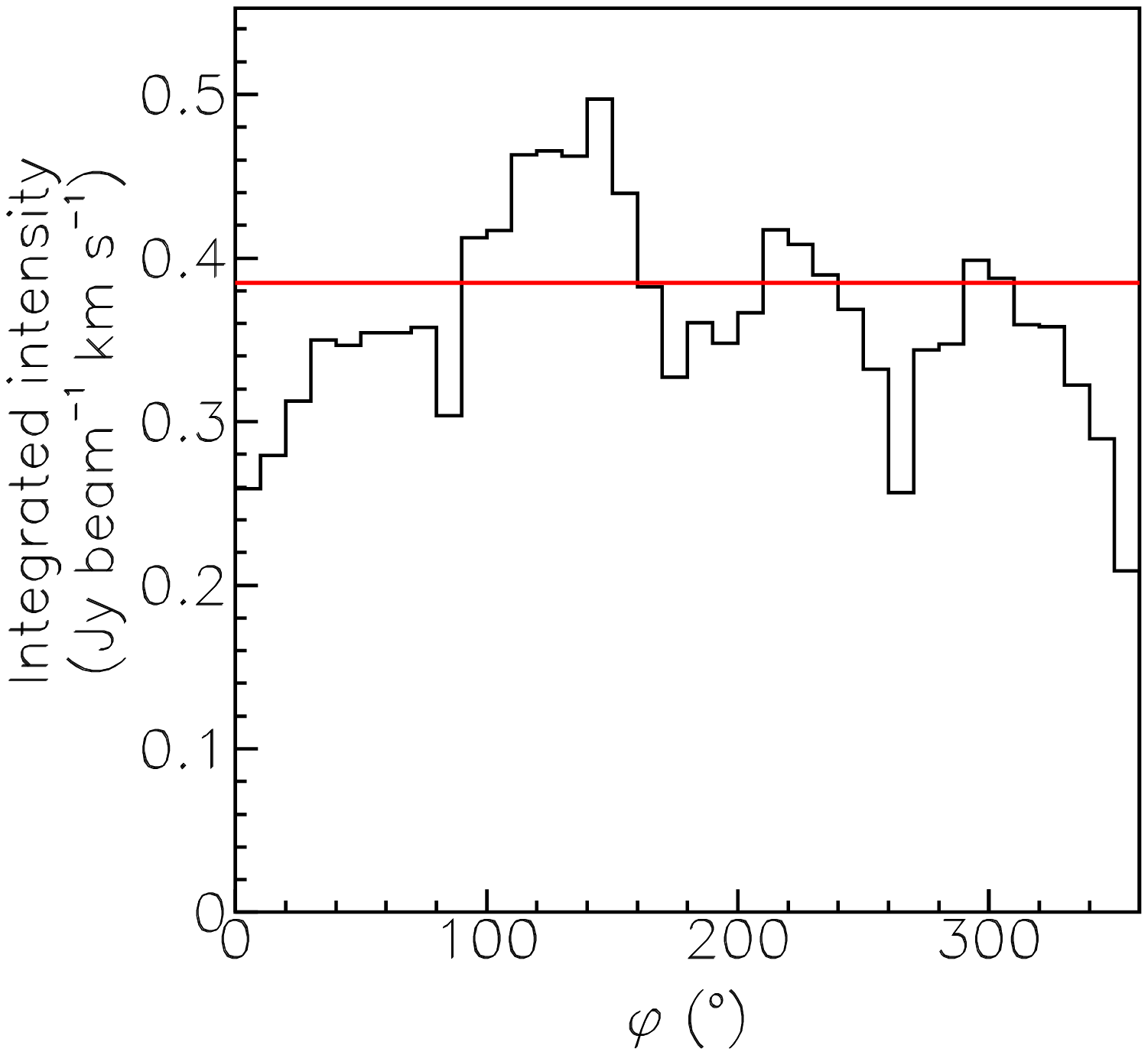}
   \caption{Line emission. {\it{Left:}} Mean value of $R$, $<R>$, weighted by the radial average of the brightness across the disc over the interval $0.54<r<2$ arcsec. 
   The red line is the result of the fit described in the text. 
{\it{Middle:}} Dependence on $\varphi$ of the mean line Doppler velocity (brightness-weighted); the red line is the result of the fit described in the text. 
{\it{Right:}} Dependence on $\varphi$ of the disc plane integrated intensity averaged across the disc ($0.54<r<2$ arcsec). The red line shows the mean value. }
   \label{Fig7}
\end{figure}

Figure~7 (left) displays the mean value of $R$, $<R>$, weighted by the radial average of the brightness across the 
ring over the interval $0.54<r<2$ arcsec. A fit of the dependence of $<R>$ on $\varphi$ as an ellipse gives 
semi-major and semi-minor axes $a_0=1.45$ arcsec and $b_0=1.19$ arcsec, position angle of the major axis $\varphi_0=97.8^\circ$ and small offsets $\Delta x=0.02$ and $\Delta y=0.07$ arcsec. 
The position angle and aspect ratio (0.82 instead of 0.85) are very similar to the dust result, but the size of the ellipse is scaled down by a factor 87\%. The tilt angle is now 35$^\circ$, compared with 32$^\circ$ for the dust.

%=======FIGURE 8 ==========
\begin{figure}
   \centering
   \includegraphics[width=13cm,trim=1.cm 1.cm 0.cm 0.cm]{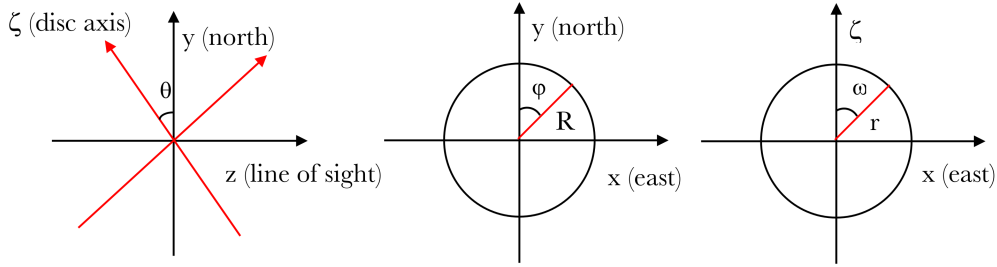}
   \caption{Geometry. {\it{Left}}: in the $(y,z)$ plane;{\it{ middle}}: in the sky plane $(x,y)$; {\it{right}}: in the disc plane $(x,\zeta)$}
   \label{Fig8}
\end{figure}

Another estimate of the tilt geometry is obtained from the map of the mean Doppler velocity (Figure~7, middle). 
In a ring defined as $0.54<r<2$ arcsec, a fit of the form $<V_z>=V_0-\Delta V \cos(\varphi-\varphi_0)$ gives $V_0=0.05$ km\,s$^{-1}$, 
$\Delta V=1.73$ km\,s$^{-1}$ and $\varphi_0=97.8^\circ$, again in excellent agreement with the value obtained from the dust fit, 
$\varphi_0=97.0^\circ$; this provides evidence against a significant in-fall contribution. The values quoted by 
Dutrey et al.~(\cite{dutrey14}) are $97^\circ\pm2^\circ$ for $^{12}$CO(6-5) and $96.5^\circ\pm0.2^\circ$ for the dust. 
The value of $\Delta V$, 1.73 km\,s$^{-1}$ corresponds to a mean rotation velocity of $\sim \Delta V/\sin\theta \sim3.3$ km\,s$^{-1}$. 
Figure~7 (right) displays the dependence on $\varphi$ of the disc plane integrated intensity averaged across the ring in the interval $0.54<r<2$ arcsec. 
It has a mean value of 0.39 Jy\,beam$^{-1}$\,km\,s$^{-1}$ and fluctuates around it with an rms of 17\%. We summarise the geometry parameters of the dust and $^{13}$CO(3-2) emission in Table~1.

%
%      TABLE1: Center positions 
%
\begin{table}
\begin{center}
\caption[]{Geometry paramters.}\label{center}

%%Please Capitalize the First Letter of Each Notional Word in table's caption

 \begin{tabular}{|c|c|c|c|c|c|c|c|c|c|c|}
  \hline
\multicolumn{2}{|c|}{} & \multicolumn{2}{|c|}{\makecell{Projection on \\$x$ and $y$}} & 
\multicolumn{6}{|c|}{Ellipse fitted to the $<R>$ vs $\varphi$} \\  
  \hline
 & & $<x>$ & $<y>$ & $a_0$ & $b_0$ & $\varphi_0$ & $\Delta\,x$ & $\Delta\,y$ & $\theta$\\
 & & (arcsec) & (arcsec) & (arcsec) & (arcsec) & ($^\circ$) & (arcsec) & (arcsec) & ($^\circ$) \\
 \hline 
\makecell{Dust} & \makecell{Central source\\Ring} & \makecell{0.06\\$-0.05$} & \makecell{$-0.13$\\$-0.09$}&
\makecell{-\\1.62} &\makecell{-\\1.38}&\makecell{-\\97.0}&\makecell{-\\$-0.07$}&\makecell{-\\$-0.05$}&\makecell{-\\32}\\ 
\hline
$^{13}$CO(3-2) & Disc & 0.02 & $-0.01$ & 1.45 &1.19 & 97.8 & 0.02 & 0.07 & 35\\
 \hline
\end{tabular}
\end{center}
\end{table}

\section{DETAILED PROPERTIES OF THE GAS DISC}
In the present section we use new coordinates obtained from those of the preceding sections by a rotation of angle $8^\circ$ about the $z$ axis. 
To within $1^\circ$, this brings the new $x$ axis on the major axes of the ellipses found in the preceding sections as best describing the $\varphi$ dependences 
of both $<R>$ and the Doppler velocity. Moreover, unless otherwise explicitly specified, we assume a tilt $\theta=35^\circ$ of the disc plane as a reasonable compromise between 
values obtained in both earlier and the present studies, for both gas and dust observations.  
In practice, we use $99\times81$ pixels of $0.06\times0.06$ arcsec$^2$ on the sky map, covering ($99\times0.06$)$\times$($81\times0.0733$)$\sim6\times6$ arcsec$^2$ 
in the disc plane ($0.0733=0.06/\cos35^\circ$). 
To each pixel ($x, y$) we associate disc coordinates $\zeta=y/\cos\theta$, $r=\sqrt{x^2+\zeta^2}$ and $\omega=90^\circ-\tan^{-1}(\zeta/x)$. Here €œdisc plane€ and €œdisc coordinates€ are 
simply defined by this transformation, implying no assumption on the disc being actually thin and flat.

\subsection{Estimate of the disc thickness obtained from the sharpness of the disc inner edge}
Tang et al.~(\cite{tang16}) have commented on the sharpness of the inner edge 
of the $^{13}$CO(3-2) emission and on the smallness of the vertical temperature gradient, 
the inner edge of the disc being directly exposed to stellar light and 
casting a shadow on the outer disc. Here, we compare the value of the smearing of the inner edge of 
the disc map near the major axis of the ellipse with its value near the minor axis. 
To a good approximation, the effect of disc thickness essentially cancels 
for the former while, for the latter, it scales with the product of the disc thickness 
by the sine of the tilt angle. The optical thickness of the line is not expected to strongly 
affect this result. We consider four angular sectors in the disc plane, 
each $60^\circ$ wide and centred on the axes of the ellipse. 
In each sector, we study the radial dependence of the integrated intensity, 
both in the disc plane ($r$) and in the sky plane ($R$). 
The result is displayed in Figure~9. 
In order to evaluate the sharpness of the inner edge of the gas disc, 
we fit a Gaussian to the rise of each distribution, 
between 0.5 and 1.5 arcsec in $r$. In $R$, we use the same interval of 0.5 to 1.5 arcsec 
for the sectors centred on the major axis of the ellipse but a scaled-down 
(by a factor $\cos35^\circ$=0.82) interval of 0.41 to 1.23 arcsec for the sectors 
centred on the minor axes in order to account for the effect of the tilt. 
The mean and $\sigma$ values (dispersions, a factor 2.35 smaller than FWHM values also commonly quoted in the literature)
obtained for the Gaussian best fits are listed in Table~2. 

%=======FIGURE 9 ==========
\begin{figure}
   \centering
   \includegraphics[width=3.72cm,trim=1cm 1.75cm 0cm 0cm]{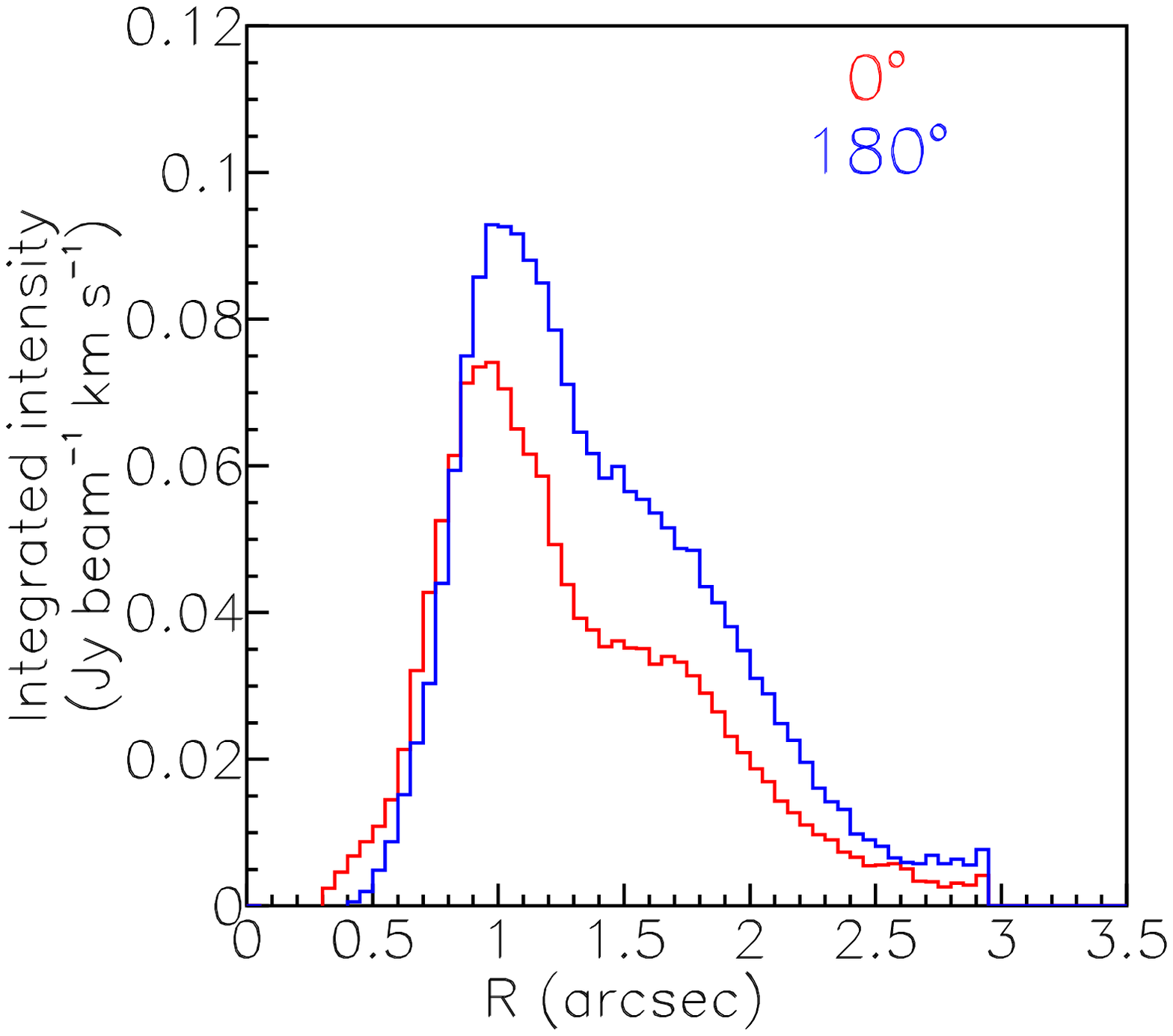}
   \includegraphics[width=3.42cm,trim=2cm 1.75cm 0cm 0cm]{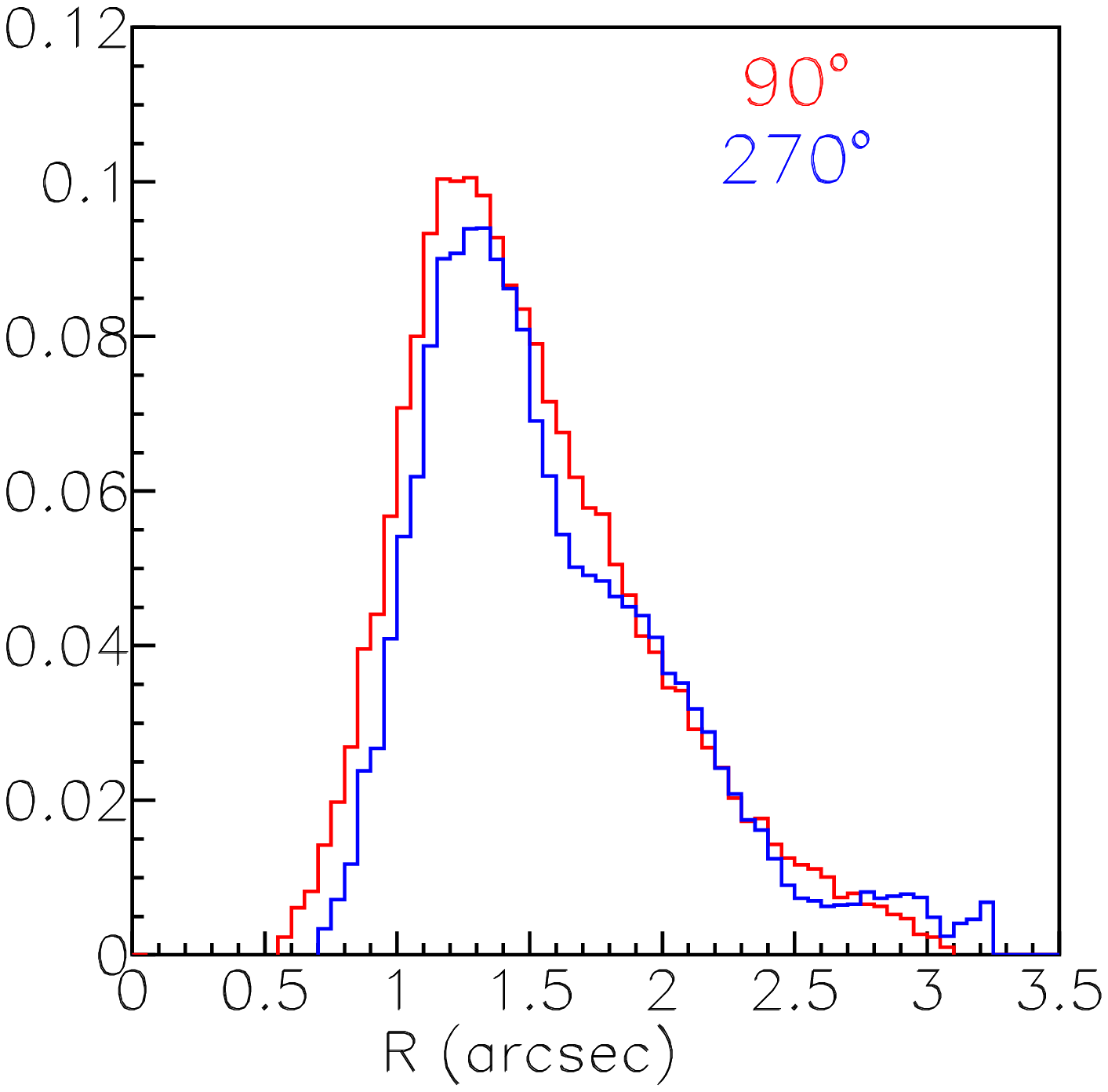}
   \includegraphics[width=3.42cm,trim=2cm 1.75cm 0cm 0cm]{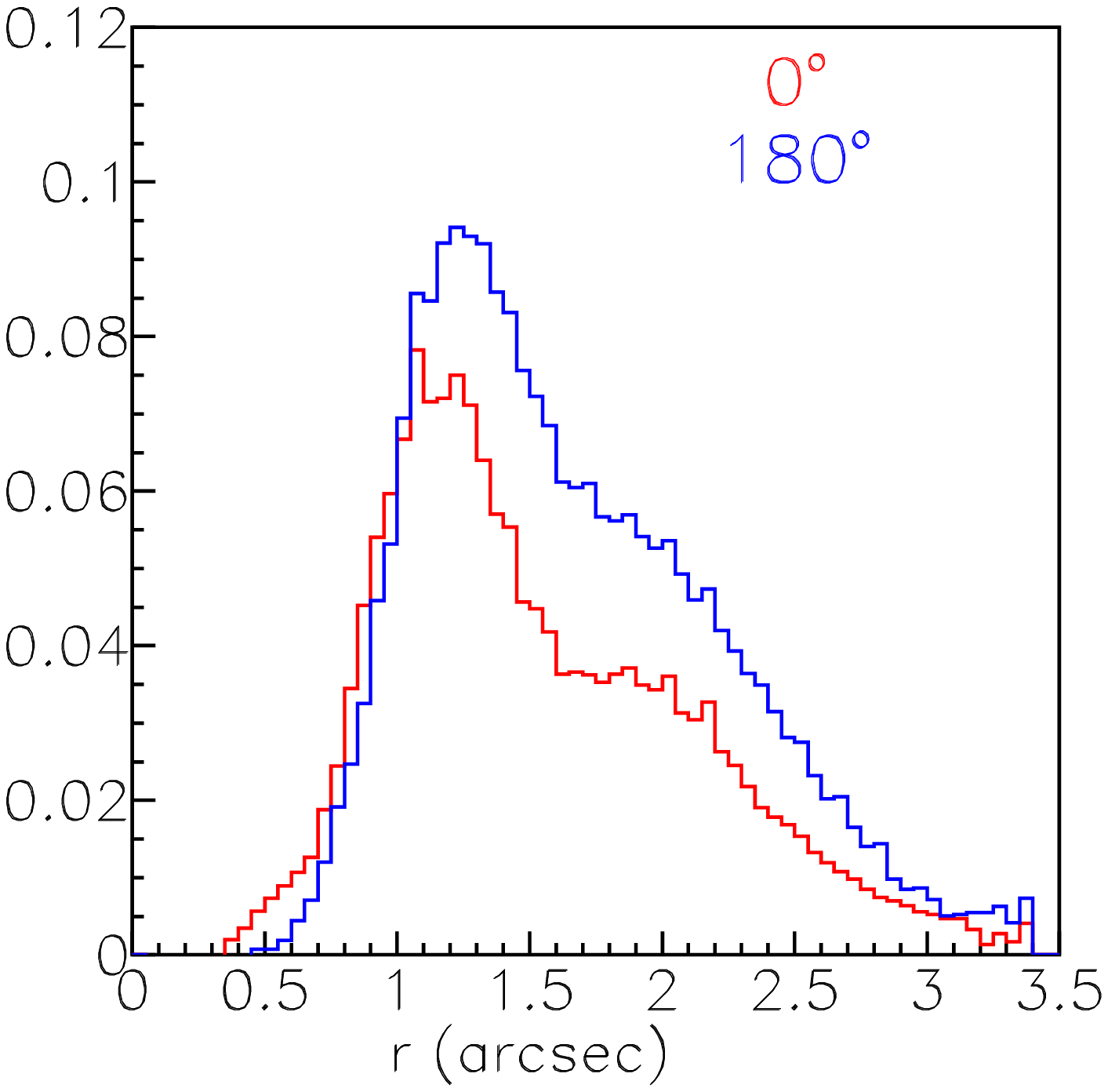}
   \includegraphics[width=3.42cm,trim=2cm 1.75cm 0cm 0cm]{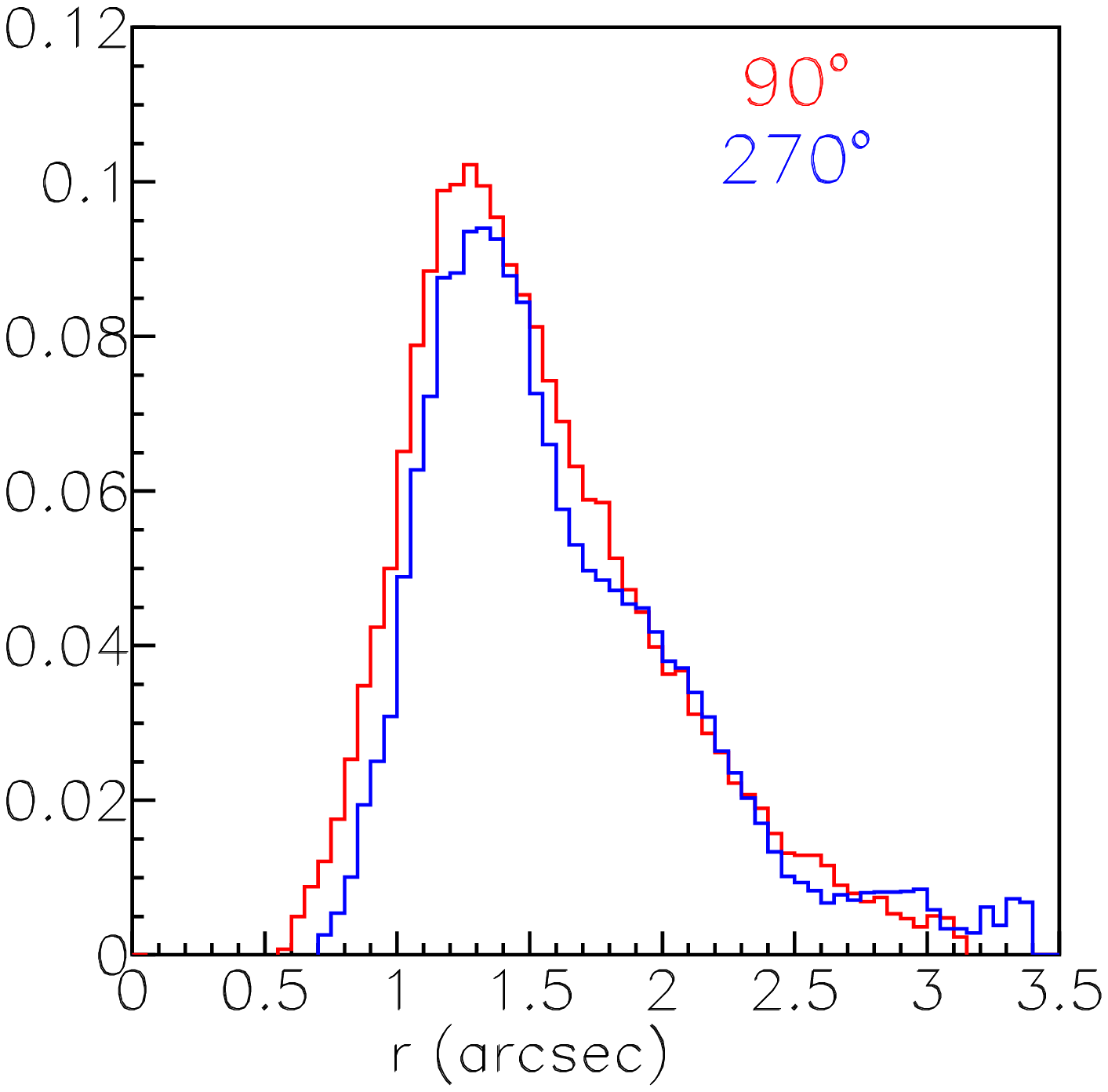}
   \caption{Line emission. Dependence on $R$ (left panels) and on $r$ (right panels) 
of the line integrated intensity averaged in $60^\circ$ wide angular sectors centred 
on the ellipse axes. In each case, the leftmost panel is for minor-axis sectors 
and the rightmost panel for major-axis sectors. 
The central values of $\omega$ are indicated in the inserts for each sector.}
   \label{Fig9}
\end{figure}
%
%      TABLE2 - Disc thickness 
%
\begin{table}
\begin{center}
\caption[]{Estimating the thickness of the gas disc from the sharpness of its inner edge projected on the sky plane. 
All values (except the scaling factors) are in arcsec.}\label{sharpness}

%%Please Capitalize the First Letter of Each Notional Word in table's caption

 \begin{tabular}{|c|c|c|c|c|c|c|c|}
  \hline
  \multicolumn{2}{ |c| }{} & North & East  & South  &  West \\  
  \hline
  $r$ fit & \makecell{$<r>$ \\ $\sigma$} & \makecell{1.15\\0.32} & \makecell{1.31\\0.34} & \makecell{1.24\\0.33} & \makecell{1.34\\0.32} \\ 
 \hline 
  $R$ fit & \makecell{$<R>$ \\$\sigma$} & \makecell{0.99\\0.28} & \makecell{1.29\\0.34} & \makecell{1.08\\0.28} & \makecell{1.32\\0.32} \\ 
\hline
  $R$ fit, beam subtracted & \makecell{$\sigma$} & 0.24 & 0.31 & 0.24 & 0.29 \\
\hline 
  $R$ fit, de-projected & \makecell{Scaling factor\\$\sigma$} & \makecell{0.86\\0.29} & \makecell{0.98\\0.31} & \makecell{0.87\\0.29} & \makecell{0.99\\0.29} \\ 
 \hline
\end{tabular}
\end{center}
\end{table}

The distributions as a function of $r$ show identical $\sigma$ values, 
to within $\pm10$ mas, in the four angular sectors. 
A contribution from the disc thickness would cause these values to be larger 
in the minor-axis sectors than in the major-axis sectors: 
it is already clear that a significant contribution from the disc thickness is excluded. 
At variance with the distributions as a function of $r$, the distributions as a function of $R$ show 
significantly different $\sigma$ values for the major-axis sectors, 
$\sim$0.33 arcsec, and the minor-axis sectors, $\sim$0.28 arcsec, a factor 85\% smaller.  
Similarly, the ratio between the mean values of the Gaussians (listed as  ``scaling factor'' in the table) 
are equal for the two sectors of a same axis of the ellipse, but again 85\% smaller for sectors 
centred on the minor axis than for those centred on the major axis. 
The latter are very slightly smaller than unity, as expected from the $60^\circ$ angular widths of the
sectors. The consistency between these numbers suggests an interpretation of the 
$\sigma$ values measured in the $R$ distributions as the sum of three terms added in quadrature: 
i) a beam contribution of 0.14 arcsec on both the minor- and major-axis sectors (calculated from the known beam parameters); 
ii) a contribution from the intrinsic smearing of the disc emission, $\sigma_0$, caused by effects such as density variations and  
contributing in each sector a value $\sigma_0$ scaled down by the scaling factors listed in the table; 
iii) an additional contribution $\sigma_1$ due to the disc thickness and contributing only 
to the minor-axis sectors. After subtraction of the beam contribution and correction 
for de-projection, one obtains values of $\sigma$ of 0.29 arcsec for the 
minor-axis sectors and $\sim$0.30 arcsec for the major-axis sectors. 
A contribution $\sigma_1$ due to the disc thickness would cause the former to exceed the latter, 
at variance with what is observed. From the consistency between the numbers, 
we estimate an uncertainty of $\sim$0.02 arcsec on the Gaussian $\sigma$s. 
To 95\% confidence level ($2\sigma$) we obtain an upper limit for $\sigma_1$ of 
$\sqrt{(0.29+2\times0.02)^2-0.30^2}=0.14$ arcsec, 
corresponding to a scale height \mbox{$H(r)\sim0.14/\sin35^\circ=0.24$ arcsec (34 au)} at 
$r\sim1$ arcsec (140 au)  where the Keplerian velocity is $\sim3$ km\,s$^{-1}$; at 30 K, 
the sound velocity is $\sim0.5$ km\,s$^{-1}$ and hydrostatic equilibrium implies 
$H(r)=0.5/3=0.17$ arcsec compared with the 0.24 arcsec upper limit obtained above. 
We have checked that this result is independent of the width of the angular sectors 
(using $40^{\circ}$ instead of $60^\circ$ lowers the Gaussian
\mbox{$\sigma$'s} by \mbox{$\sim 0.01$ arcsec}). 
Depending on the interval chosen to calculate the Gaussian \mbox{$\sigma$'s} lower values of the $\sigma_1$ 
upper limit may be obtained, as low as 0.10 arcsec instead of 0.14 arcsec. 
We conservatively prefer to retain the latter value as our final result. 

\subsection{Integrated intensity variations across the disc}
In order to better understand the nature of the integrated intensity variations 
displayed in Figure~7 (right), we display in Figure~10 the map in the disc plane 
of the difference between the measured integrated intensity and its value averaged 
over $\omega$ at the same value of $r$ (as obtained from Figure~6, right). 
This map provides a measure of the lack of rotational symmetry of the 
integrated intensity in the disc plane. It gives strong evidence for an excess 
associated with the hot spot observed by Dutrey et al.~(\cite{dutrey14}) and Tang et al.~(\cite{tang16}) 
and for a northern depression of similar amplitude. Both excess and depression 
reach their maxima at a distance from the central stars corresponding to 
the gap€ between the maxima of the two first Gaussians describing the mean 
radial distribution of the integrated intensity (these Gaussians peak at $r$=1.22 and 1.87 arcsec respectively). 
It is also in this gap€ that the continuum dust emission peaks (at $r$=1.62 arcsec). 
However, both excess and depression extend to larger values of $r$, 
particularly the former that extends out to $r\sim2.5$ arcsec.
%    FIGURE 10 
\begin{figure}
   \centering
   \includegraphics[width=5.cm,trim=1.cm 1cm 2cm 0cm]{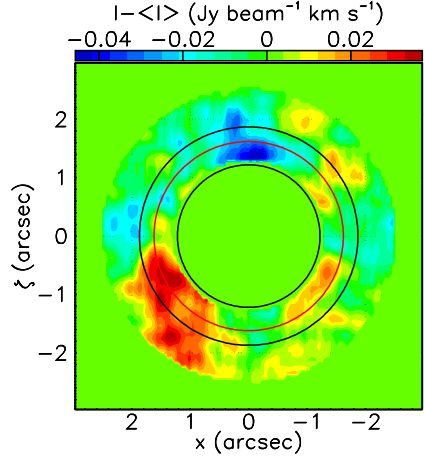}
   \caption{Line emission. Map in the disc plane of the difference between the integrated intensity 
and its value averaged over $\omega$ at the same $r$. 
The black circles show the maxima of the Gaussians describing the mean 
radial integrated intensity distribution, $r$=1.22 and 1.87 arcsec respectively. 
The red circle corresponds to the mean value of $r$ in the dust map (1.62 arcsec).}
   \label{Fig10}
\end{figure}

\subsection{Gas kinematics}
Calling $V_{rot}$ and $V_{fall}$ the components of the disc plane velocity 
respectively perpendicular and parallel to the disc radius, 
the Doppler velocity reads $V_z=\sin\theta$($V_{rot}\sin\omega-V_{fall}\cos\omega$) 
for each data-cube element ($x,y,V_z$). To a good approximation, 
$V_{fall}$ can be neglected and we can calculate $V_{rot}=V_z(\sin\theta\sin\omega)^{-1}$ 
for each data-cube element, leaving for later the task to reveal a possible small $V_{fall}$ contribution. 
$V_{rot}$ becomes trivially singular along the $\zeta$ axis. 
We require accordingly $|\sin\omega|$ to exceed 0.3 when calculating $V_{rot}$. 
As $\sin^{-1}(0.3)=17.5^\circ$, this is not much of a loss.

Figure~11 displays the dependence on $\omega$ and $r$ of $V_z$ averaged 
(using brightness as weight) over \mbox{$0.8<r<2.5$} arcsec and over $\omega$ respectively. 
Averaging $V_z$  requires some care in dealing with the noise: 
the interval used for averaging must be symmetric with respect to the mean value obtained as a result, 
which requires relaxing the condition $|V_z|<2$ km\,s$^{-1}$ usually applied in the analysis. 
The $\omega$-dependence is perfectly described by a sine wave of amplitude $-1.43$ km\,s$^{-1}$. 
Adding a cosine term does not change the coefficient of the sine term and 
insignificantly improves the value of $\chi^2$. 
Its amplitude is $-0.05$ km\,s$^{-1}$, only 2.6\% of the amplitude of the $\sin\omega$ term, 
corresponding to a shift of $-1.9^\circ$ in $\omega$. 
As a check of the correctness of the procedure, we compare this result with 
what is obtained when requiring a 3-$\sigma$ cut on each data-cube element; 
the amplitudes of the sine and cosine waves become $-1.40$ and $-0.04$ km\,s$^{-1}$ respectively. 

The negative sign of the best-fit cosine term means radial expansion, 
in-fall would give a positive sign. Assuming a 3$^\circ$ uncertainty on $\omega$,
corresponding to half a beam sigma at a distance of 1.3 arcsec,
we obtain a 3-$\sigma$ upper limit (99\% confidence level) of 9\% on the ratio 
$V_{fall}/V_{rot}$. As a function of $r$, averaging over $\omega$ would cause $<V_z>$ 
to cancel if symmetry with respect to the $\zeta$ axis were perfect. 
It is indeed found very small, at the level of $-0.05$ km\,s$^{-1}$ 
as soon as $r$ exceeds the peak of the radial integrated intensity distribution 
at $r\sim1.3$ arcsec.
 
Similarly, the dependence on $\omega$ and $r$ of $V_{rot}$ averaged respectively 
(using brightness as weight) over $0.8<r<2.5$ arcsec and over $\omega$ ($|\sin\omega|>0.3$) 
is displayed in Figure~12. The left panel shows the distribution of 
$<V_{rot}\,r^{1/2}>$ on $r$, which would be constant if the rotation were Keplerian. 
A fit in the interval $1.1<r<2.5$ arcsec gives a power index of $-0.63$ instead of 
the Keplerian $-0.5$ and \mbox{$<V_{rot}>=3.0$ km\,s$^{-1}$} at $r=1$ arcsec. 
The middle panel illustrates the difficulty to measure $V_{rot}$ reliably due to 
its singularity on the $\zeta$ axis. As remarked earlier, 
the binarity of the central star prevents the position of the "centre" 
to be defined to better than some $\pm0.1$ arcsec 
(more exactly such a definition requires modelling properly the binary configuration). 
Shifting the origins of $x$ and $y$ on the sky map by $\pm1$ pixel size 
($\pm0.06$ arcsec) changes the value of $\omega$ and therefore of $V_{rot}$. 
The result displayed in the middle panel shows the importance of the effect. 
As a result, increasing the $|\sin\omega|$ cut from 0.3 to 0.707 
($\sin45^\circ$) makes the $<V_{rot}\,r^{1/2}>$ distribution Keplerian 
with a power index of $-0.51$ instead of $-0.63$, 
the rotation velocity at $r=1$ arcsec increasing from 3.0 to 3.1 km\,s$^{-1}$ (left panel). 
We show in the right panel the map of $V_{rot}\,r^{1/2}$ in the disc plane. 
It is uniform except for increases near the $\omega$ limits 
in the north-west and south-east directions. These are largely artefacts due to the difficulty 
of calculating reliably $V_{rot}$ near the $\zeta$ axis. 
Note that Dutrey et al.~(\cite{dutrey14}) quote a $V_{rot}$ value of 
$3.4\pm0.1$ km\,s$^{-1}$ for $^{12}$CO(6-5) emission with an index of 
$-0.5\pm0.1$ at $r=100$ au; this corresponds to 2.9 km\,s$^{-1}$ at $r=1$ arcsec, 
consistent with the \mbox{3.0 km\,s$^{-1}$} observed here for $^{13}$CO(3-2) emission.
%    FIGURE 11 
\begin{figure}
   \centering
   \includegraphics[width=5.2cm,trim=0cm 0cm 0.5cm 2cm]{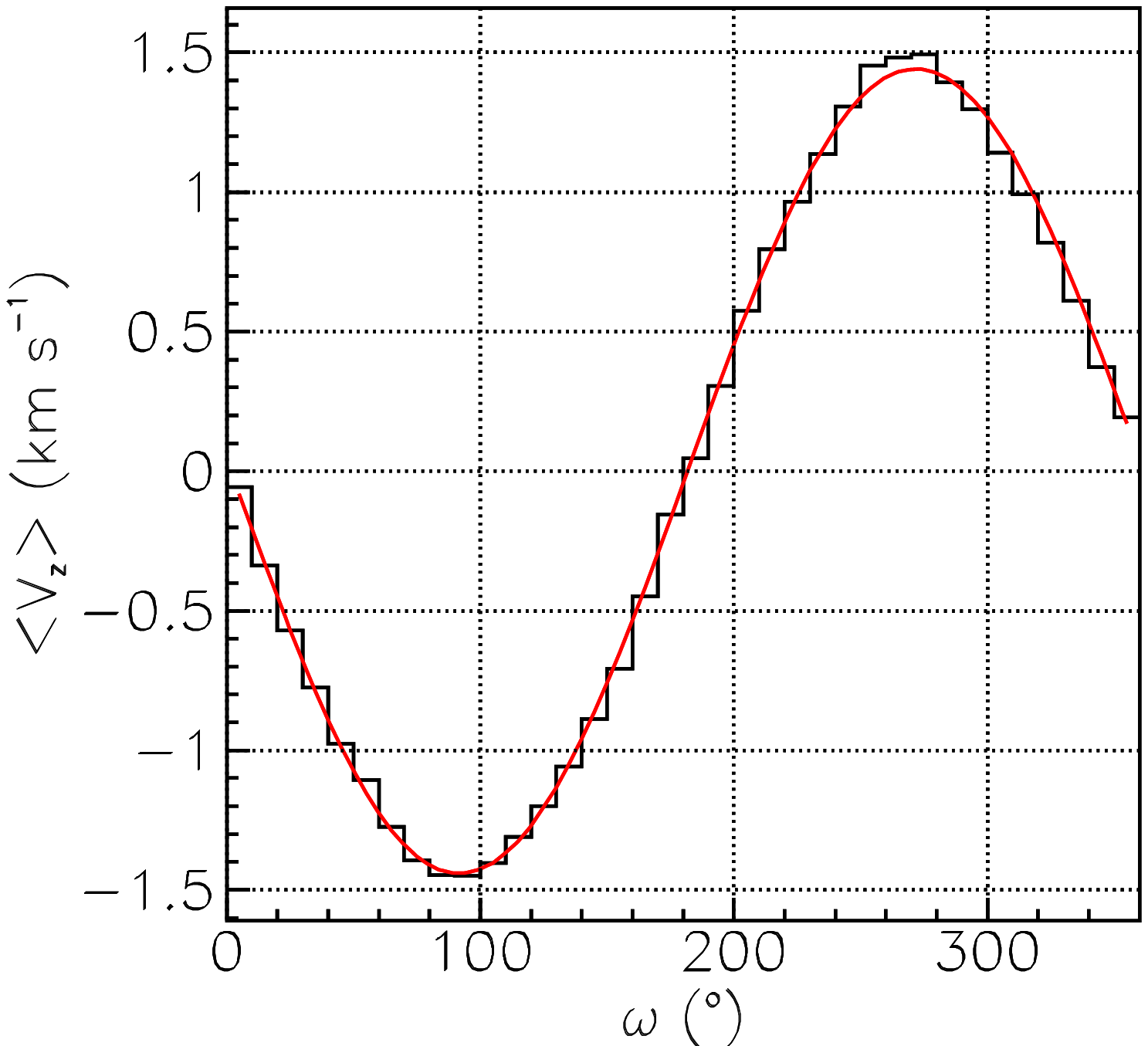}
   \includegraphics[width=5.2cm,trim=0cm 0cm 0.5cm 2cm]{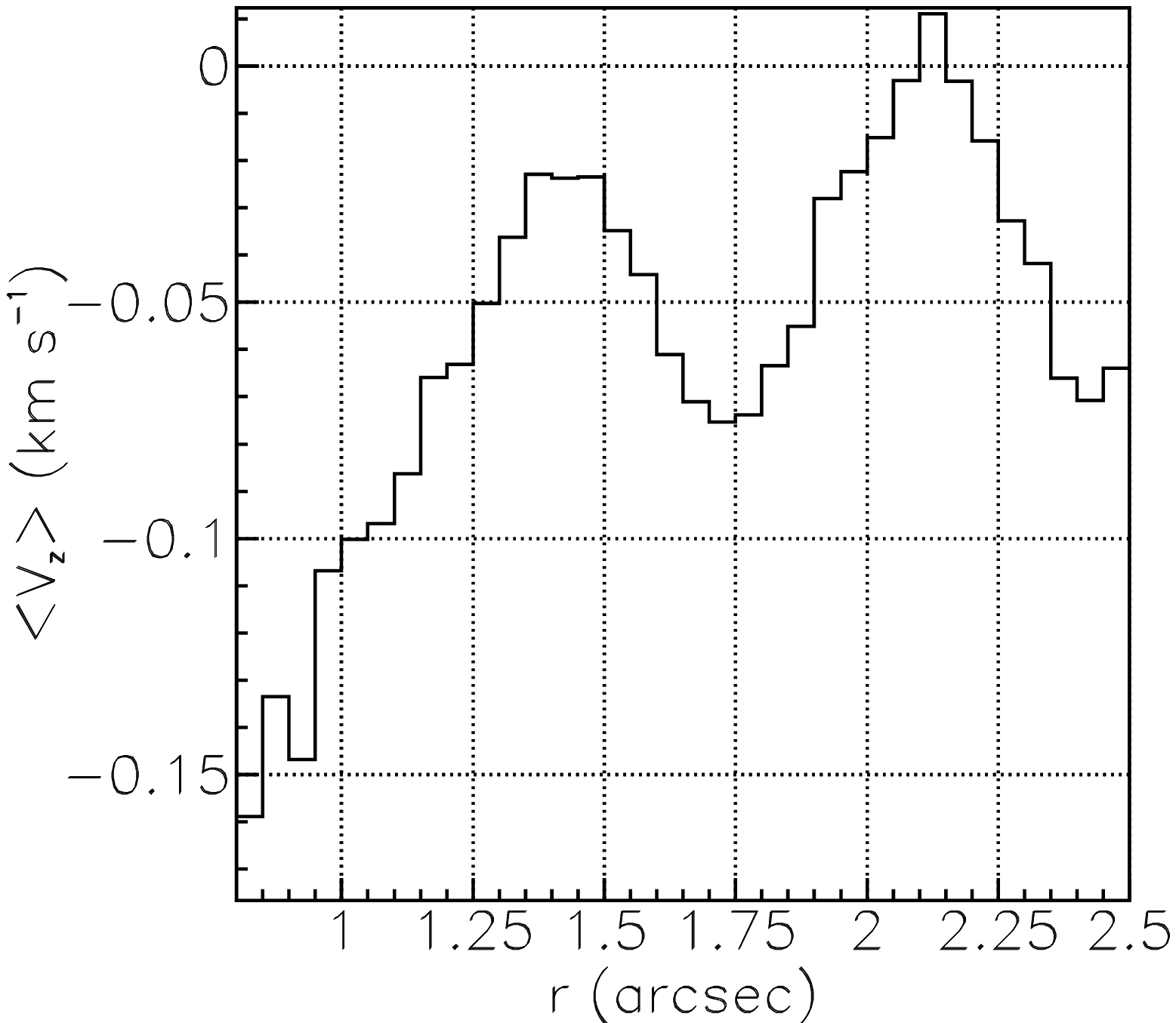}
   \caption{Distributions on $\omega$ (left) and $r$ (right) of the Doppler velocity 
respectively averaged over $0.8<r<3.2$ arcsec and over $\omega$. 
In the left panel, the line shows the best fit result, of the form 
\mbox{$-1.43\sin\omega+0.05\cos\omega=-1.43\sin(\omega-1.9^\circ)$ km\,s$^{-1}$}.}
   \label{Fig11}
\end{figure}
% FIGURE 12 
\begin{figure}
   \centering
   \includegraphics[width=4.65cm,trim=1cm 0cm 0cm 1cm]{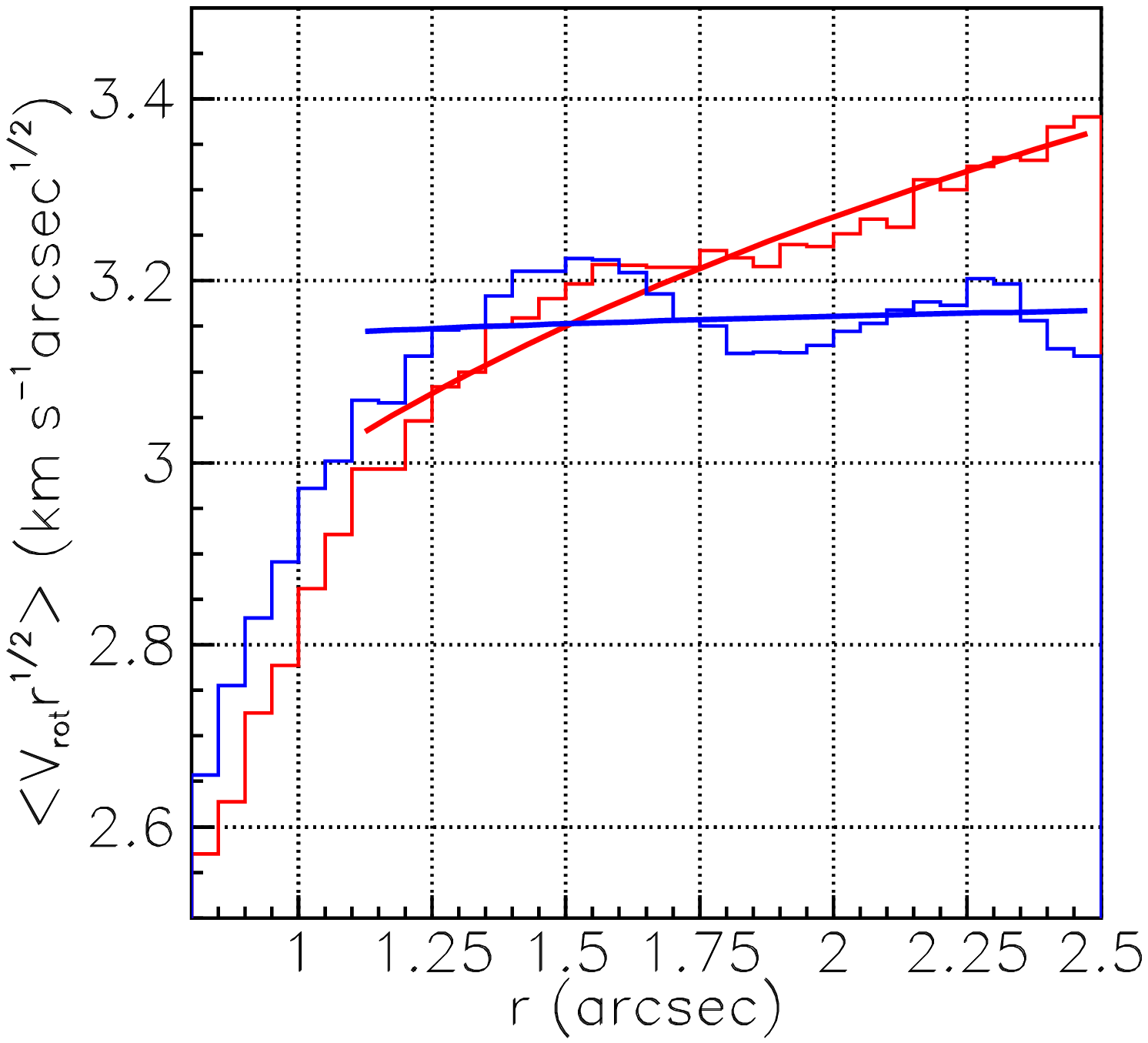}
   \includegraphics[width=4.65cm,trim=1cm 0cm 0cm 1cm]{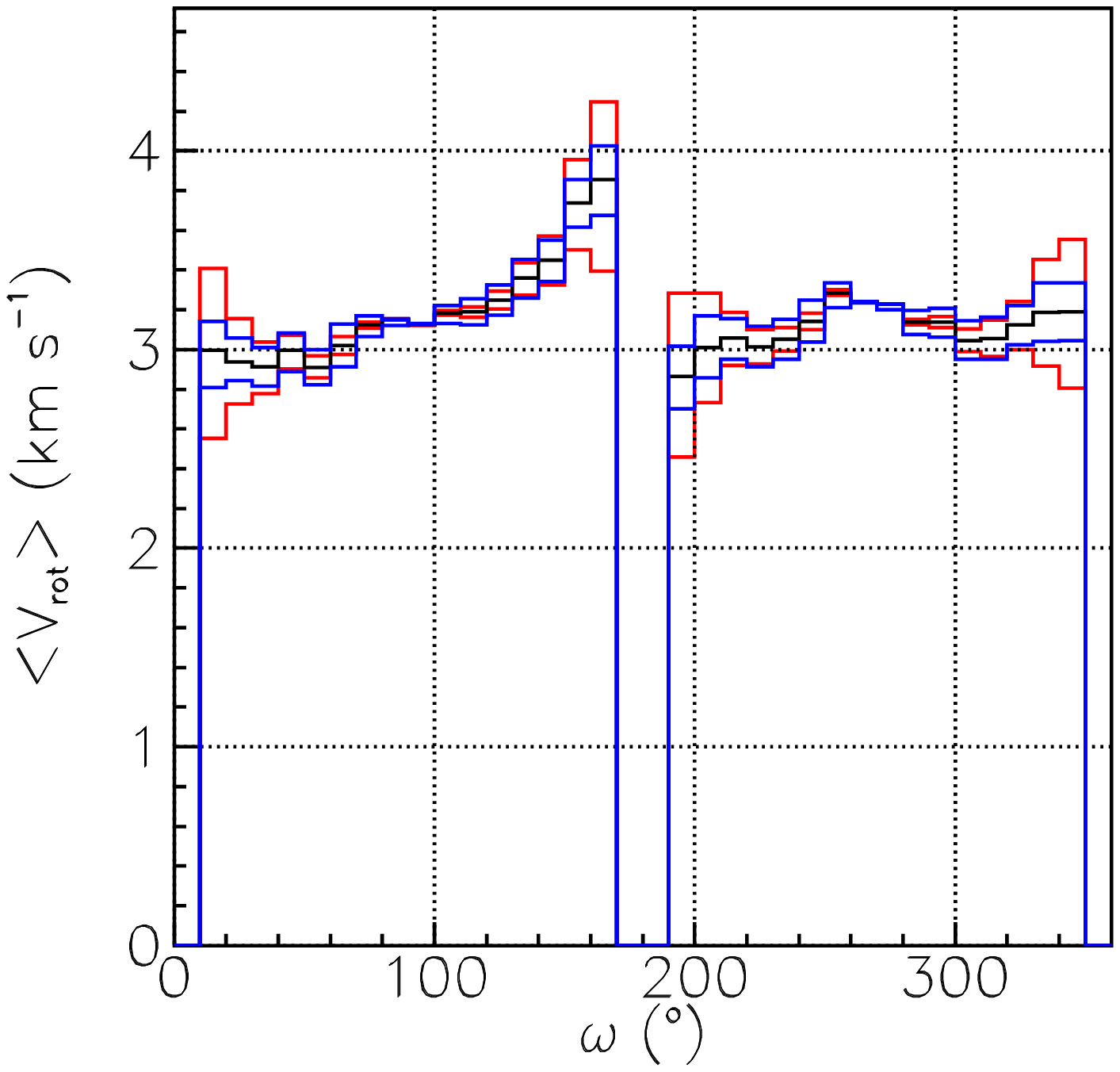}
   \includegraphics[width=4.65cm,trim=1cm 0cm 0cm 1cm]{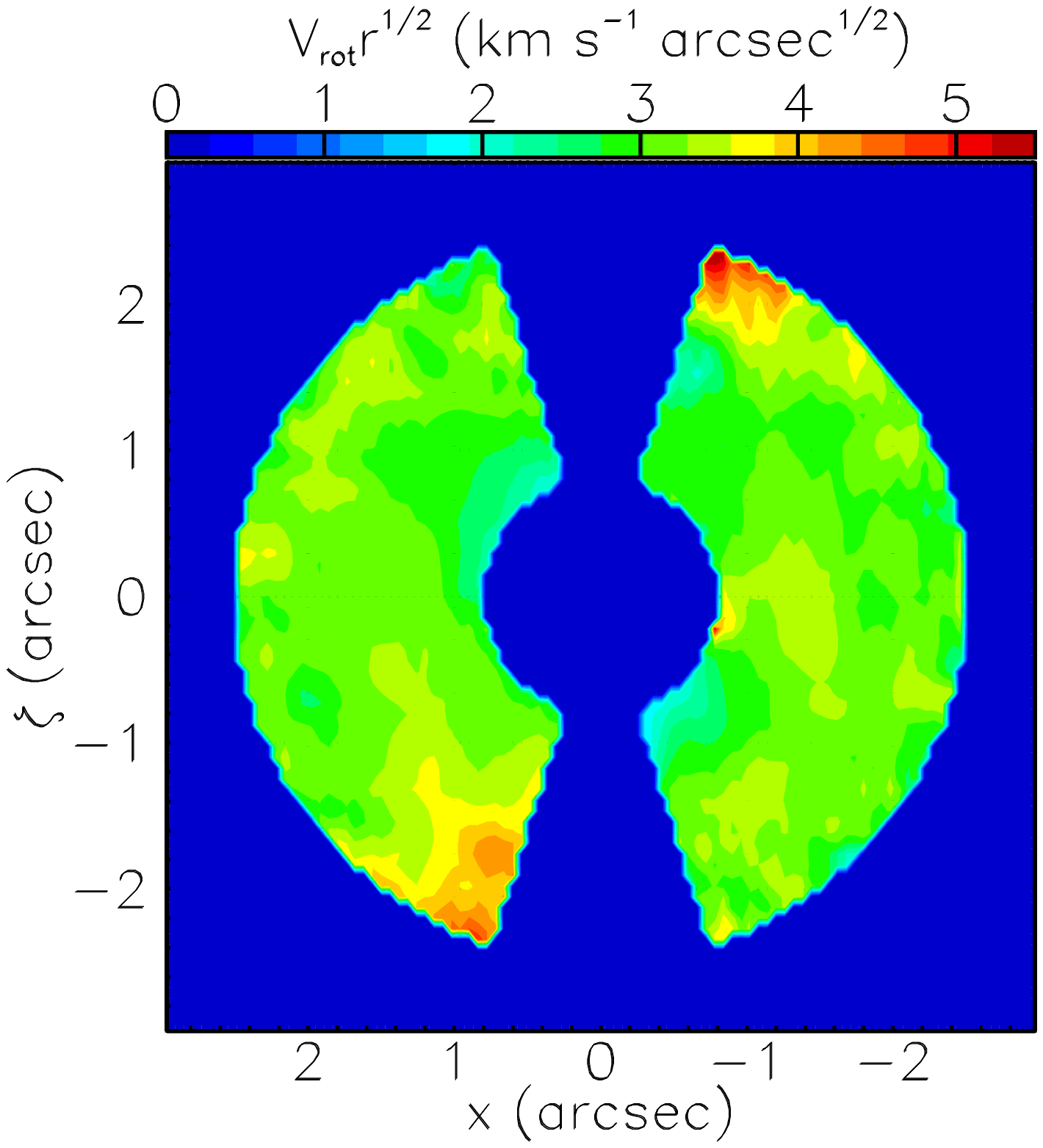}
   \caption{{\it{Left:}} Dependence on $r$ of $<V_{rot}\,r^{1/2}>$ (brightness-weighted average); 
the lines are the best power law fits with indices $-0.63$ for $|\sin\omega|>0.3$ (red) and 
$-0.51$ for $|\sin\omega|>0.707$ (blue). 
{\it{Middle:}} Dependence on $\omega$ of $<V_{rot}>$ (averaged in the interval $0.8<r<2.5$) 
calculated using the nominal origin of coordinates on the sky plane 
(black histogram) or by shifting the origin by $\pm0.06$ arcsec in either $x$ or $y$ 
(red and blue histograms). {\it{Right:}} De-projected map of $<V_{rot}\,r^{1/2}>$ ($|\sin\omega|>0.3$).}
   \label{Fig12}
\end{figure}

\subsection{Line width}
Figure~13 displays the dependence of the brightness on the difference $dV_z$  
between the values of $V_z$ measured in a given pixel and their mean values in that same pixel.
The mean is calculated using brightness as a weight and the histogram is summed over all pixels in the 
interval \mbox{$0.8<r<2.5$ arcsec}. A Gaussian fit gives a $\sigma$-value of \mbox{0.23 km\,s$^{-1}$}. 
  
Several quantities, added in quadrature, are expected to contribute to $\sigma_{vz}$: 
Keplerian shear $\sigma_K$ associated with both beam size and disc thickness, 
the instrumental resolution $\sigma_I$ and the thermal broadening $\sigma_T$, 
possibly including a turbulence contribution (Teague et al., ~\cite{Teague2016}), and opacity broadening, $\sigma_{\tau}$. 

Averaged over $\omega$, the Keplerian shear at $r=1.5$ arcsec reads, 
from the derivative of a power law, \mbox{$\sigma_K=0.6<|V_z|>\sigma_r/r$} 
where 0.6 stands for the power index of the radial $V_{rot}$ distribution (it would be 0.5 in a pure Keplerian case). 
Here, $\sigma_r$ is the sum in quadrature of the $\sigma$'s of beam, \mbox{$0.21$ arcsec}, and of the disc thickness
multiplied by \mbox{$\tan\theta=0.7$}, $0.11$ arcsec. Hence, \mbox{$\sigma_r\sim\sqrt{0.21^2+0.11^2}=0.24$ arcsec} and 
\mbox{$\sigma_K=0.6\times\sin(35^\circ)\times3.1\times1.5^{-0.6}\times(2/\pi)\times\sigma_r/1.5=0.09$ km\,s$^{-1}$}.
Taking the FWHM of the instrumental resolution as one velocity bin gives \mbox{$\sigma_I\,\sim0.05$ km\,s$^{-1}$}.
Thermal broadening proper reads $\sqrt{2kT/M_{co}}$ where $k$ is Boltzmann constant, 
$T$ the temperature and $M_{co}$ the mass of the $^{13}$CO molecule; 
at $T=18$\,K (Tang et al.~\cite{tang16}) it amounts again to some \mbox{0.10 km\,s$^{-1}$}. 
Opacity tends to flatten the line profile and its effect is an effective broadening of $\sim\sqrt{ln(\tau)}$, where $\tau$ 
is the line opacity (Pi\'etu et al.~\cite{pietu2007}). Tang et al.,~(\cite{tang16}) show that 
$^{13}$CO(3-2) and $^{12}$CO(3-2) have similar brightness, implying that $\tau$($^{13}$CO) is significantly above unity.
Using both $^{13}$CO(3-2) and $^{12}$CO(3-2), we estimate its value to be $\tau\sim10$, 
meaning an effective broadening of $\sim$1.5 and a joint contribution of $\sim0.15$ km\,s$^{-1}$ 
for thermal and opacity broadening. A possible additional source of broadening could be the effect of noise. 
However, using a 3-$\sigma$ cut to select the data, 
which must underestimate the measured value of $\sigma_{vz}$, 
we obtain 0.20 instead of 0.23 km\,s$^{-1}$, showing that noise can be neglected within our estimated uncertainty of $\pm0.03$ km\,s$^{-1}$.

Adding the estimated contributions in quadrature gives a total contribution of 
\mbox{$\sqrt{0.05^2+0.09^2+0.15^2}=0.18$ km\,s$^{-1}$} compared with \mbox{$0.23\pm0.03$ km\,s$^{-1}$} measured: there is not much room left for additional contributions and turbulence is small (highly subsonic) in this disc. 

Important additional information on the line width can be obtained 
from a study of the variations of $\sigma_{vz}$ over the disc plane. 
To this end we consider three $r$ intervals, 0.4 arcsec wide, 
covering between 1.3 and \mbox{2.5 arcsec} and 24 $\omega$-intervals, $15^\circ$ wide, 
covering between 0 and 360$^\circ$. 
The dependence on $\omega$ of the integrated intensity is shown in Figure~14 (left) 
for each $r$-interval separately. The hot spot sticks out at values of $\omega$ 
that increase from $\sim120^\circ$ in the low $r$-interval to $\sim150^\circ$ in the high $r$-interval. 
The middle panel shows the dependence on $\omega$ of the normalized value of $V_z$ averaged in each $r$-$\omega$ bin separately; 
more precisely a fit of the form $<V_z>=-a\,\sin\omega-b\cos\omega$ is performed in each $r$-$\omega$ bin separately 
and the normalization is made by dividing each of the three distributions by the corresponding value of $a$ 
(respectively 1.46, 1.27 and 1.18 km\,s$^{-1}$, namely $\sim1.78$ km\,s$^{-1}$ divided by $<r>^{1/2}$). 
The values of $b$ are between 0.02 and 0.03 km\,s$^{-1}$ and can be neglected. 
All three normalized histograms are well described by a sine wave. 
The right panel displays the dependence on $\omega$ of $\sigma_{vz}$: in each $r$-$\omega$ interval the $\sigma$ 
of a Gaussian fit to the peak of the Doppler velocity spectrum is plotted after normalization 
to its value averaged over $\omega$ in the corresponding $r$-interval (0.258, 0.210, and 0.181 km\,s$^{-1}$ respectively).
%FIGURE 13
\begin{figure}
   \centering
   \includegraphics[width=5.25cm,trim=1cm 0.75cm 0cm 1cm]{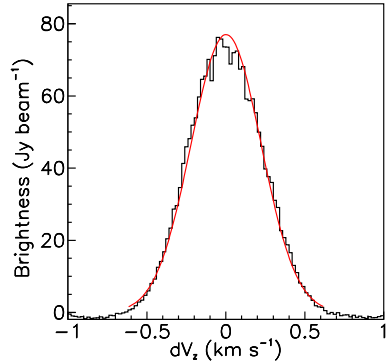}
   \caption{Dependence of the brightness on the difference $dV_z$ between measured values of $V_z$ their mean values in the pixel. Summing is over all pixels in the interval $0.8<r<2.5$ arcsec. The curve is a Gaussian fit.}
   \label{Fig13}
\end{figure}

As a function of $\omega$, the line width fluctuates relatively 
less than the integrated intensity. Moreover, there is no sign of 
a sine wave contribution that would signal the effect of Keplerian shear, 
confirming the conclusion that was reached above. 
While the hot spot dominates the variations of the integrated intensity, 
its presence is barely visible as an increase of the line width; 
conversely, sharper line width excesses at $\omega\sim60^\circ$ and $320^\circ$ 
are visible on the velocity-integrated distribution as less marked excesses. 
The depressions at $\omega\sim0^\circ$, $90^\circ$ and $270^\circ$ 
are also associated with lower values of the line width. 
The correlation between $\sigma_{vz}$ and fluctuations of the integrated intensity $f$ is 
illustrated in Figure~15. In each ($r, \omega$) bin we define 
$\Delta\sigma$ and $\Delta f$ as the difference between the values of 
$\sigma_{vz}$ and $f$ and their mean in the $r$ interval: 
$\Delta\sigma=\sigma_{vz}/<\sigma_{vz}>-1$ and $\Delta f=f/<f>-1$. 
A clear positive correlation is evidenced from the best linear fit, 
$\Delta\sigma=0.32\,\Delta f$. Note that the correlation is even 
slightly stronger if one excludes the hot spot region, the corresponding Pearson coefficients being respectively 0.25 and 0.32.
From the low-$r$ interval to the high-$r$ interval the $\omega$-averaged line width ($\sigma$) 
decreases by a factor 0.70 while the amplitude of the $V_z$ sine wave 
decreases only by a factor 0.81.
A possible explanation may be an increase
of the temperature and opacity with decreasing $r$. 
An increase of temperature and opacity from \mbox{($T, \tau$)=(18\,K, 5)} at \mbox{$r\sim2.3$ arcsec} 
to (36\,K, 10) at $r\sim1.5$ arcsec would imply an effective thermal broadening
increasing from $\sim0.13$ km\,s$^{-1}$ to about 0.21 km\,s$^{-1}$. 
Adding in quadrature $\sigma_K$ and $\sigma_I$ contributions of 
respectively 0.09 and 0.05 km\,s$^{-1}$ would give respectively 0.15 and 0.23 km\,s$^{-1}$,
compared with 0.18 and 0.26 km\,s$^{-1}$ being measured.

\begin{figure}
   \centering
   \includegraphics[width=4.4cm,trim=0.5cm 1.25cm 0.5cm 0.2cm]{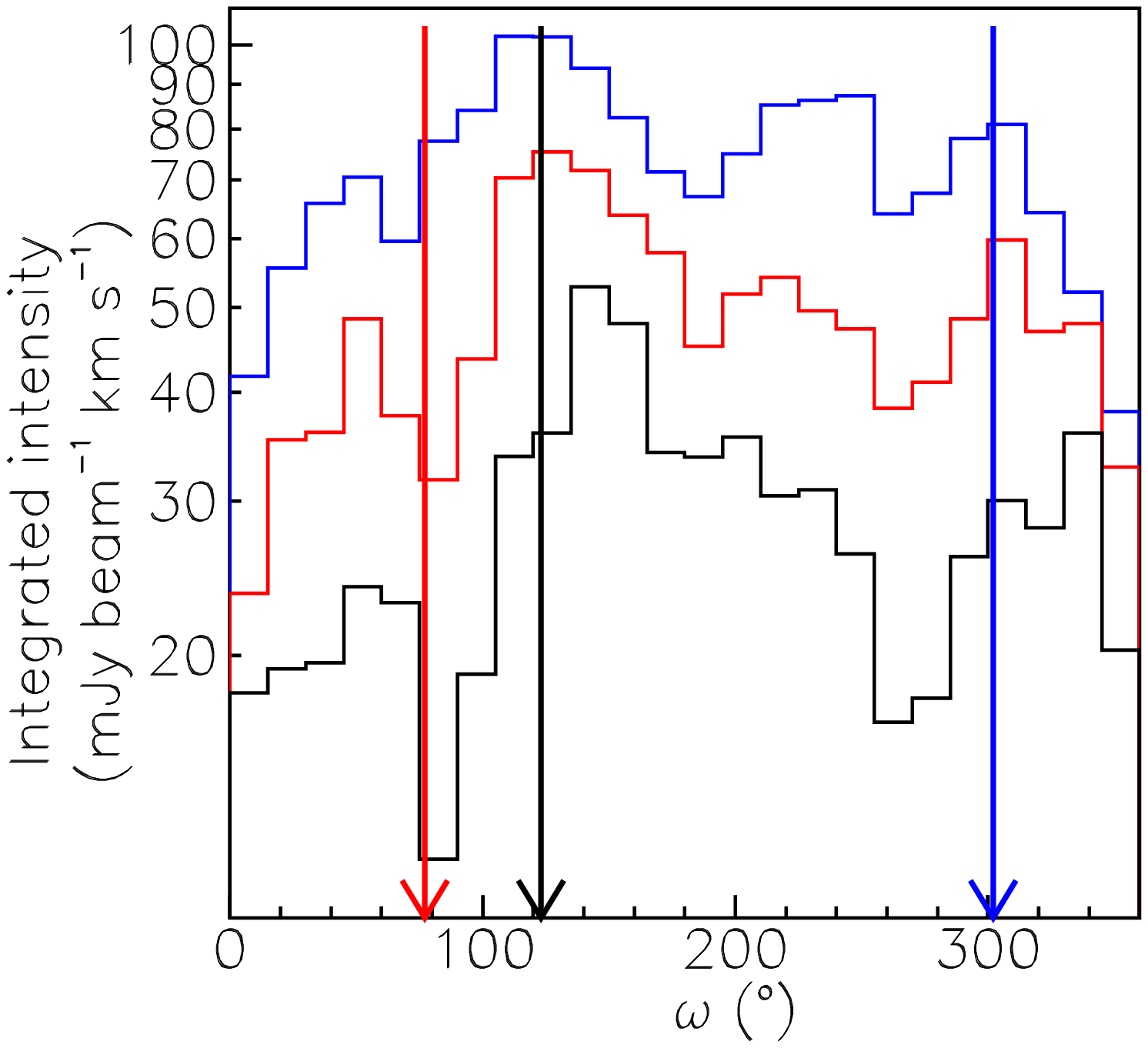}
   \includegraphics[width=4.4cm,trim=0.5cm 1.25cm 0.5cm 0.2cm]{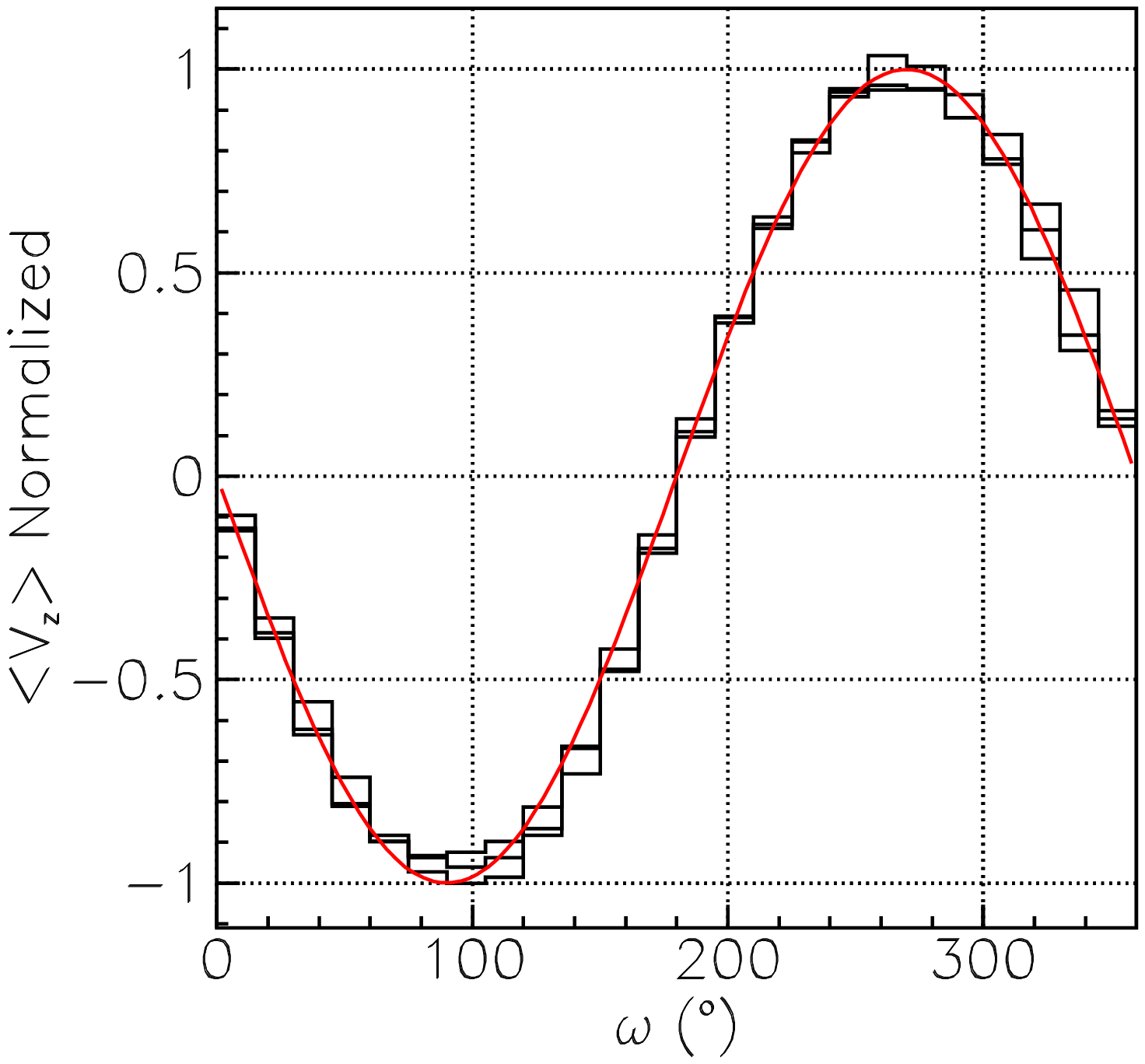}
   \includegraphics[width=4.4cm,trim=0.5cm 1.25cm 0.5cm 0.2cm]{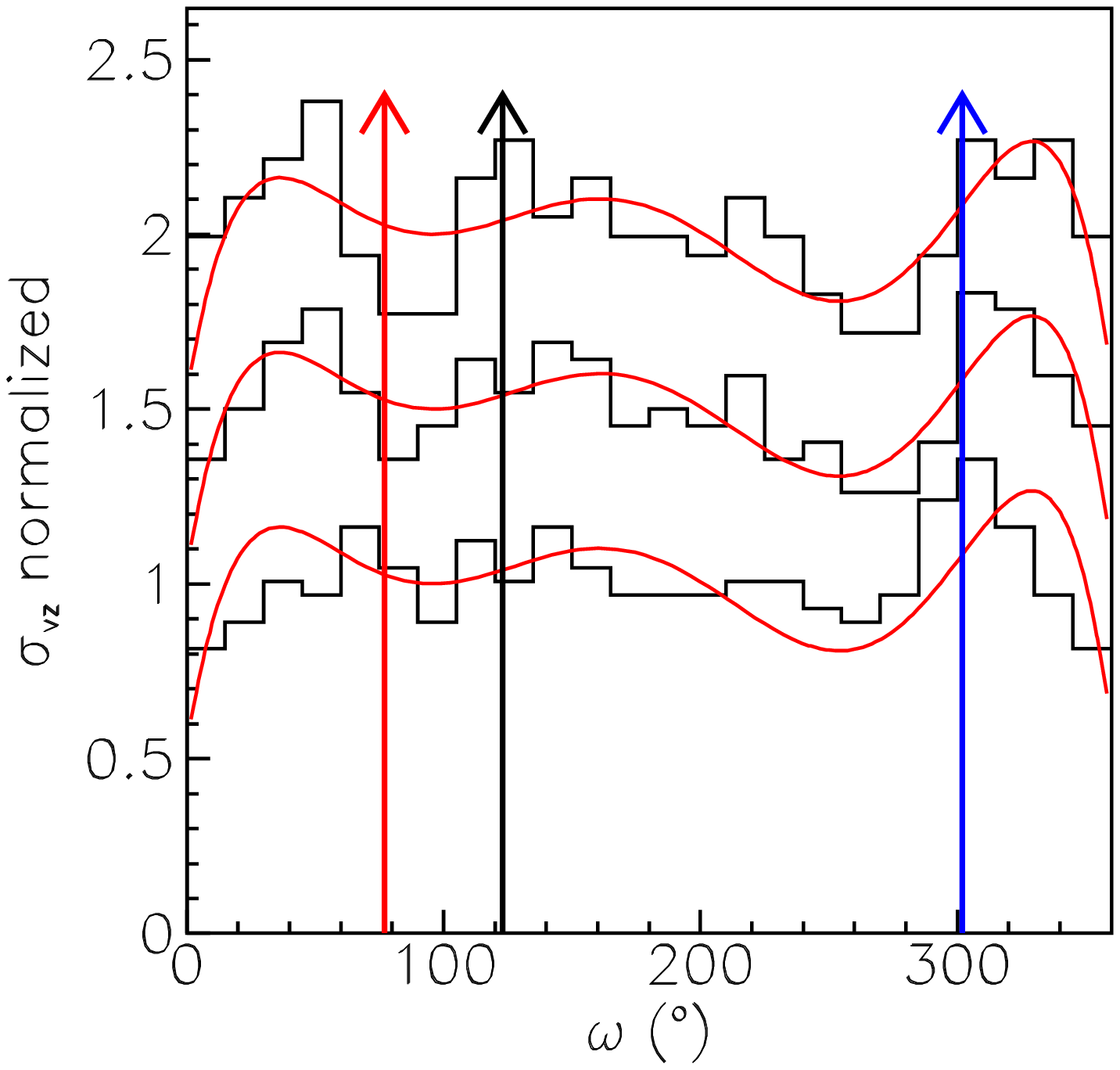}
   \caption{{\it{Left:}} Dependence on $\omega$ of the integrated intensity 
for \mbox{$1.3<r<1.7$ arcsec (blue)}, \mbox{$1.7<r<2.1$ arcsec (red)} and \mbox{$2.1<r<2.5$ arcsec (black)}; 
{\it{Middle:}} dependence on $\omega$ of the value of $<V_z>$ in each of the 
three $r$-intervals (black histograms); 
here, $<V_z>$ has been divided by 1.46, 1.27 and 1.18 km\,s$^{-1}$ respectively, 
making the three histograms nearly identical; the red curve is a sine wave. 
{\it{Right:}} dependence of $\sigma_{vz}$ on $\omega$, for each $r$-interval separately; 
in each $r$-$\omega$ bin, a Gaussian fit is performed to the peak of the $V_z$ spectrum, 
giving a $\sigma$-value that averages to respectively 0.258, 0.210 and 0.181 km\,s$^{-1}$; 
the plotted histograms are normalized to these respective average values; 
in addition, for clarity, they are shifted up by respectively 0, 0.5 and 1. 
The red curve, a sixth degree polynomial fit to the distribution of the central $r$-interval, 
is shown to guide the eye. In the left and right panels the arrows point in the 
direction of increasing $r$ and indicate remarkable features: 
the black arrow shows the hot spot as defined from the left panel, 
the blue and red arrows show peaks of the line width as defined from the right panel.}
   \label{Fig14}
\end{figure}
% FIGURE 15
\begin{figure}
   \centering
   \includegraphics[width=4.8cm,trim=1.5cm 2cm 2cm 1cm]{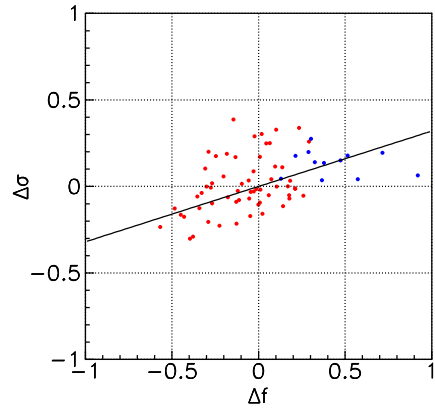}   
   \caption{Correlation between the normalized relative fluctuations of the line width 
$\Delta\,\sigma$ and the integrated intensity $\Delta\,f$ (see text). 
The line is the best fit to the data, $\Delta\,\sigma=0.32\Delta\,f$. 
The blue points are for $105^\circ<\omega<165^\circ$ (hot spot).}
   \label{Fig15}
\end{figure}

The fact that the $\omega$-dependence of $<V_z>$ is very well described 
by a simple sine wave in each of the three intervals implies that 
the observations are consistent with $V_{rot}$ being independent of $\omega$. 
It shows again that the fluctuations of $V_{rot}\,r^{1/2}$ observed in Figure~11 
are affected by very large uncertainties. Indeed, very good fits to 
the measured $V_z$ sky map are obtained by assuming a purely Keplerian rotation velocity. 
On the contrary, we estimate that the uncertainty attached to both $<V_z>$ and $\sigma_{vz}$ 
is of the order of only 0.02 km\,s$^{-1}$, making the discussion 
of the line width in terms of $\sigma_{vz}$ more reliable than in terms of $V_{rot}$ dispersion.

\section{Summary}
\label{sect:discussion}
In summary, the present analysis confirms the results obtained earlier by 
Tang et al.~(\cite{tang16}). It reveals the presence of concentric 
dust and gas rings sharing a same axis projecting on the 
sky plane $\sim7^\circ$ east of north. 
In the approximation where both rings are flat and thin, 
their inclination angles with respect to the sky plane 
are respectively $32^\circ$ and $35^\circ$. 
The gas ring is broader than the dust ring and peaks at 
smaller distance (typically 87\%) to the central stars. 
The de-projected radial dependence of the line emission displays maxima 
at $\sim1.2$ and 1.9 arcsec from the central stars, 
bracketing the mean dust ring radius of \mbox{$\sim1.6$ arcsec}. 
It cuts-off sharply at a mean distance of $\sim1$ arcsec, cancelling completely 
below \mbox{$\sim0.54$ arcsec}. Azimuthal rms variations of the dust and gas emissions 
in the disc planes are measured at the respective levels of 
$\sim\,\pm9\%$ and $\pm17\%$. 
Strong evidence is obtained for the rotation of the tilted gas disc 
about its axis dominating the kinematics.

A detailed study of the properties of the gas disc has been presented, 
adding significant new contributions to the earlier analysis of 
Tang et al.~(\cite{tang16}). From the azimuthal dependence of the sharpness 
of the inner edge of the disc, a 95\% confidence level upper limit of 
0.24\,arcsec (34\,au) has been placed on its scale height $H(r)$ at 
a distance of 1\,arcsec (140\,au) from the central stars. At 30\,K, 
hydrostatic equilibrium would imply $H(r)/r\sim0.17$, 
consistent with this observation.  

Variations of the integrated intensity across the disc area have been 
studied in detail and found to confirm the presence of a hot spot 
in the south-eastern quadrant. However several other significant fluctuations, 
in particular a depression in the northern direction, have also been revealed. 
On average, the rms relative variation of the integrated intensity reaches 
only $\sim17\%$. The radial dependence of the integrated intensity is 
modulated with enhancements at $r\sim1.2$ and 1.9 arcsec, 
bracketing the dust ring ($\sim1.6$ arcsec). 
It is also between these radial integrated intensity enhancements 
that both the hot spot and the northern depression are observed 
to peak (their effects nearly cancelling each other when averaged over $\omega$).
 
The study of the gas kinematics has given evidence for a strong dominance 
of rotation about the disc axis. The Doppler velocity gradient being perpendicular 
to the projection of the disc axis on the sky plane allows placing a 99\% confidence 
upper limit of 9\% on the ratio between a possible in-fall velocity and the rotation velocity. 
The difficulty of evaluating reliably the rotation velocity close to the sky plane projection 
of the disc axis has been illustrated and commented upon. 
Taking this in proper account, the rotation is observed to be Keplerian 
with a power index of $\sim-0.51$ across most of the disc area. 
At $r=1$ arcsec, the rotation velocity reaches $\sim3.1$ km\,s$^{-1}$ 
in agreement with the value measured by Dutrey et al.~(\cite{dutrey14}) 
for $^{12}$CO(6-5). No significant anomaly can be revealed in 
regions of important integrated intensity variation 
such as the hot spot and the northern depression.
 
Finally, the dependence of the line width on $r$ and $\omega$ 
has been studied. It shows little dependence on $\omega$ and 
increases as $r$ decreases: the $\sigma$ of the line, $\sigma_{vz}$, 
increases from $\sim0.18$ km\,s$^{-1}$ to $\sim0.26$ km\,s$^{-1}$ 
when $r$ decreases from 2.3 to 1.5 arcsec. As the contributions of 
Keplerian shear and instrumental spectral resolution taken together 
should not exceed some 0.11 km\,s$^{-1}$, a possible explanation may be a factor $\sim2$ decrease 
of the disc surface temperature and opacity, reaching respectively 36\,K and 10 at \mbox{$r=1.5$ arcsec}.
Relative variations of the line width over the disc area 
have been found to be strongly correlated with relative variations 
of the integrated intensity, the former being about a third of the latter. 
At least qualitatively, this result would also support the presence of a temperature gradient, 
the CO(3-2) emission ladder peaking at temperatures higher than the average disc temperature.

These new results contribute significant additional information 
and complement the earlier conclusions reached by Dutrey et al.~(\cite{dutrey14}) 
and Tang et al.~(\cite{tang16}). However, considerations on optical thickness, 
which are discussed in detail by Tang et al.~(\cite{tang16}), 
are not repeated here. Moreover, interpretations of the observed variations 
of the integrated intensity as signalling the formation of a 
planet or of a new companion star remain valid suggestions 
that would require detailed modelling to be validated. 
However, this is beyond the scope of the present work.

\begin{acknowledgements}
This paper makes use of the following ALMA data: ADS/JAO.ALMA\#2012.1.00129.S. 
ALMA is a partnership of ESO (representing its member states), NSF (USA) and NINS (Japan), 
together with NRC (Canada), NSC and ASIAA (Taiwan), and KASI (Republic of Korea), 
in cooperation with the Republic of Chile. 
The Joint ALMA Observatory is operated by ESO, AUI/NRAO and NAOJ. 
This research has made use of the SIMBAD database, operated at CDS, Strasbourg, France, and of the NASA ADS Abstract Services.
We thank the anonymous referee for useful comments that helped improving the presentation of this work.
We thank the members of the initial ``GG Tau team'' for early contributions, 
Jeff Bary, Tracy Beck, Herv\'e Beust, Yann Boehler, Frederic Gueth, 
Jean-Marc Hur\'e, Vincent Pi\'etu, Arnaud Pi\'erens and Michal Simon. 
This research is funded by Vietnam National Foundation for Science and
Technology Development (NAFOSTED) under grant number 103.99-2016.50.
Department of Astrophysics (VNSC/VAST) acknowledges support from the World Laboratory, Rencontres du Viet Nam, the Odon Vallet fellowships, Vietnam National Space Center and Graduate University of Science and Technology.  
Anne Dutrey and St\'ephane Guilloteau thank the French CNRS programs PNP, PNPS and PCMI. 

\end{acknowledgements}

\appendix                  %%appendicial material is supported

\label{lastpage}

\end{document}